\documentclass[
journal=mamobx,
manuscript=article,
layout=twocolumn,
email=true
]{achemso}

\usepackage{mathtools}
\usepackage{amssymb}
\usepackage{graphicx} % include figure files
\usepackage{dcolumn}  % align table columns on decimal point
\usepackage[utf8]{inputenc}
\usepackage[
colorlinks=true,
linkcolor=blue,
citecolor=blue,
urlcolor=blue]{hyperref} % add hypertext capabilities
\usepackage{color}
\usepackage{xr}
\setkeys{acs}{doi=true, articletitle=true}
\SectionNumbersOn
\newcommand\numberthis{\addtocounter{equation}{1}\tag{\theequation}}
\newcommand{\doi}[1]{\href{http://dx.doi.org/#1}{\nolinkurl{#1}}}
\title{Particle Diffusivity and Free-Energy Profiles in Inhomogeneous Hydrogel Systems from Time-Resolved Penetration Profiles}
\author{Amanuel Wolde-Kidan}
\altaffiliation{These two authors contributed equally to this work.}
\affiliation{Fachbereich Physik, Freie Universität Berlin, Arnimallee 14, 14195 Berlin, Germany}
\author{Anna Herrmann}
\altaffiliation{These two authors contributed equally to this work.}
\affiliation{Institut für Chemie und Biochemie, Freie Universität Berlin, Takustr. 3, 14195 Berlin, Germany}
\author{Albert Prause}
\author{Michael Gradzielski}
\affiliation{Institut für Chemie, Straße des 17. Juni 124, Technische Universität Berlin, 10623 Berlin, Germany}
\author{Rainer Haag}
\author{Stephan Block}
\affiliation{Institut für Chemie und Biochemie, Freie Universität Berlin, Takustr. 3, 14195 Berlin, Germany}
\author{Roland R. Netz}
\email{rnetz@physik.fu-berlin.de}
\affiliation{Fachbereich Physik, Freie Universität Berlin, Arnimallee 14, 14195 Berlin, Germany}
\begin{tocentry}
  \begin{center}
    \includegraphics{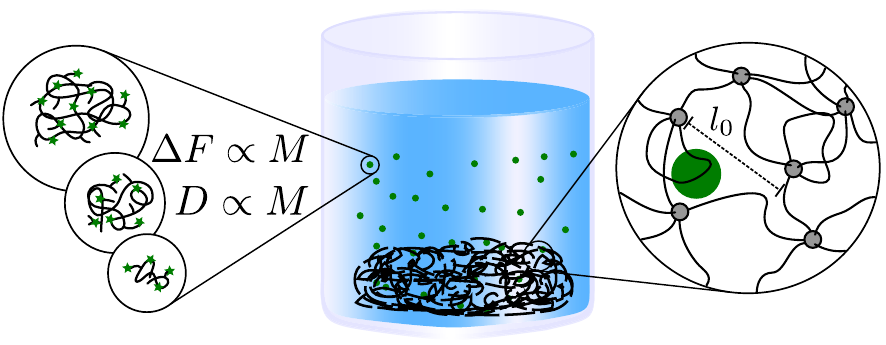}
  \end{center}
\end{tocentry}
\begin{document}

\begin{abstract}
A combined experimental/theoretical method to simultaneously determine diffusivity and free-energy profiles of particles that penetrate into inhomogeneous hydrogel systems is presented. As the only input, arbitrarily normalized concentration profiles from fluorescence intensity data of labeled tracer particles for different penetration times are needed. The method is applied to dextran molecules of varying size which penetrate into hydrogels of polyethylene-glycol (PEG) chains with different lengths that are covalently cross-linked by hyperbranched polyglycerol (hPG) hubs. Extracted dextran bulk diffusivities agree well with fluorescence correlation spectroscopy data obtained separately. Scaling laws for dextran diffusivities and free energies inside the hydrogel are identified as a function of the dextran mass. An elastic free-volume model that includes dextran as well as PEG linker flexibility describes the repulsive dextran-hydrogel interaction free energy, which is of steric origin, quantitatively and furthermore suggests that the hydrogel mesh-size distribution is rather broad and particle penetration is dominated by large hydrogel pores. Particle penetration into hydrogels is for steric particle-hydrogel interactions thus suggested to be governed by an elastic size-filtering mechanism that involves the tail of the hydrogel pore-size distribution.
\end{abstract}
\clearpage

\section{Introduction}\noindent
The penetration of particles into hydrogels is relevant for
technological applications~\cite{Wirthl2017, Herrmann2018}, drug delivery~\cite{Li2016} and for biological systems such as biofilms~\cite{Billings2015b}, the extracellular matrix~\cite{Rosales2016} and mucus~\cite{Lieleg2010}. Mucus, which is the most common biological hydrogel, lines the epithelial tissues of different organs, such as the respiratory, gastrointestinal and urogenital tracts.
Mucus is mainly composed of mucins, which are glycoproteins of varying length that absorb large amounts of water and thereby lend mucus its hydrogel nature, and additional components such as enzymes and ions~\cite{Wagner2018}. Mucins are relevant in the cell signaling context and presumably also play a role in the development of cancer~\cite{Hollingsworth2004}. But primarily, mucus is a penetration barrier against pathogens, e.g. virions or bacteria, while it allows the permeation of many non-pathogens, e.g. nutrients, that are absorbed through the mucosa of the small intestine~\cite{McGuckin2011}.
Studies have suggested that, based on the type of mucus, different mechanisms give rise to the protective barrier function~\cite{Dawson2003, Lai2007}, in addition to the advective transport of pathogens through mucus shedding or clearance~\cite{Johansson2012, Button2012}, which is not considered here.
One typically distinguishes steric size-filtering mechanisms from interaction-filtering mechanisms~\cite{Lieleg2010, Witten2017}, the latter presumably play a major role in the defense of organisms against pathogens since they allow for precise regulation of the passage of wanted and unwanted particles and molecules~\cite{Li2013, Marczynski2018}. Recent studies demonstrated that attractive electrostatic interactions reduce particle diffusivity inside hydrogels substantially and much more than repulsive electrostatic interactions~\cite{Zhang2015, Hansing2016} and that salt concentration and the distribution of charges and pore size are important parameters which influence the permeation properties of charged hydrogels~\cite{Hansing2018a, Hansing2018b}.

Particle penetration into mucus and biofilms has been studied by single-particle tracking techniques~\cite{Dawson, Wagner2017} as well as by methods where a diffusor ensemble is observed~\cite{Wilking2013, Li2013, Lawrence1994, Marczynski2018}.
On short time scales, transient particle binding to the hydrogel~\cite{Zhang2015, Hansing2016, Marczynski2018} is important and leads to anomalous particle diffusion~\cite{Cherstvy2019}. On spatial length scales larger than the hydrogel mesh size and on time scales larger than typical  binding escape times, particle diffusion is in a continuum description determined by the free-energy and diffusivity profiles across an inhomogeneous hydrogel system.
In this framework, particle binding effectively reduces the diffusivity.
If the free-energy and diffusivity profiles are known, particle penetration can be quantitatively predicted, provided the particle concentration is low and the particles do not modify the hydrogel properties in an irreversible manner. In this context it should be noted that both profiles depend on the interactions between particle and hydrogel and therefore are different for each distinct hydrogel-particle pair. Due to method restrictions, experiments primarily focussed on determining either the particle diffusivity inside the hydrogel~\cite{Lieleg2010, Dawson2003, Dawson} or on the partitioning between hydrogel and the bulk solution~\cite{Sassi1996}, from which the free energy inside the hydrogel (relative to the bulk solution) can be determined. However, for prediction of the penetration or permeation speed of particles into the hydrogel, both the diffusivity and the free energy in the hydrogel are needed.

\begin{figure*}
\includegraphics[width=\textwidth]{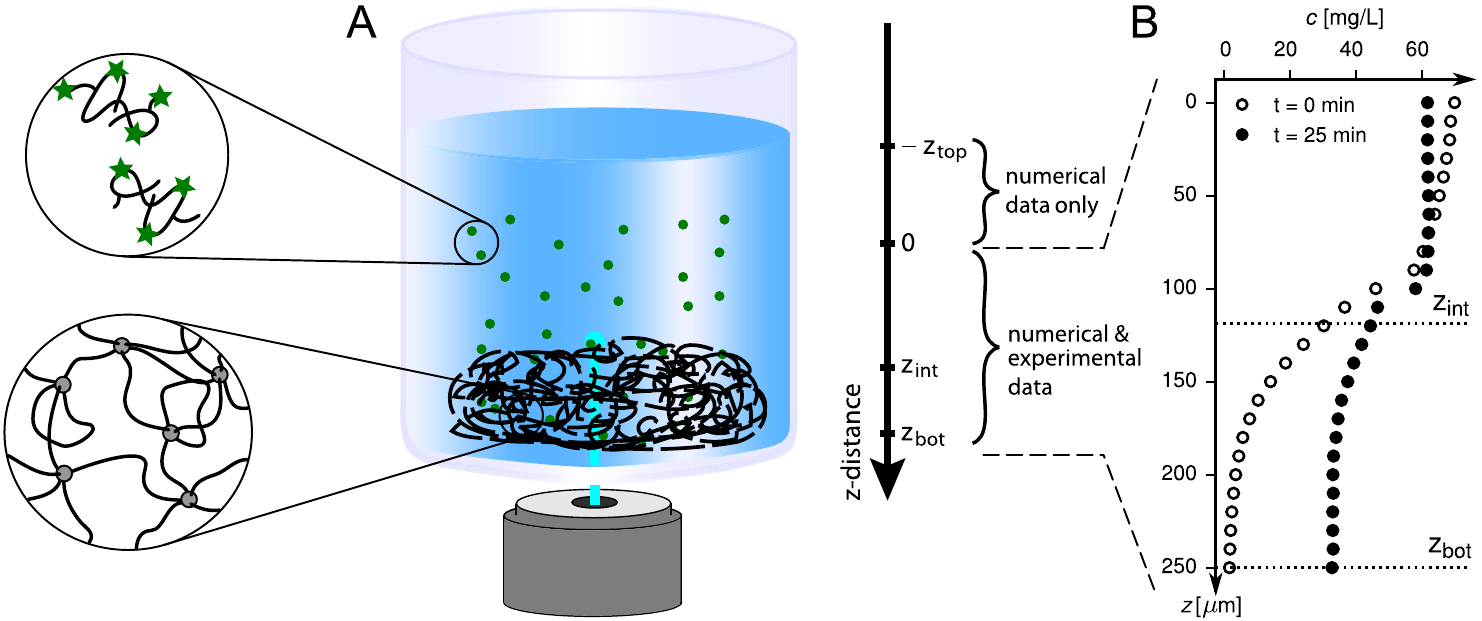}
\caption{A:~Schematic drawing of the experimental setup. Concentration profiles of fluorescently labeled dextran molecules (green) are measured as they penetrate from the bulk solution (blue) into the hydrogel (black). The origin of the z-axis is positioned such that experimentally measured profiles range from $z = 0$ to $z = z_{\text{bot}}$. The hydrogel-bulk solution interface is located at $z = z_{\text{int}}$. In the range from $z = -z_{\text{top}}$ to $z = 0$ only numerically determined concentration profiles are available.
B:~Exemplary experimental concentration profiles for two different penetration times for $M_\text{dex}=4~$kDa dextran diffusing into the \textit{hPG-G10} hydrogel, positions of the hydrogel-bulk solution interface $z_{\text{int}}$ and the hydrogel-glass bottom interface $z_{\text{bot}}$ are indicated.}
\label{fig:introduction}
\end{figure*}

In this work, we study synthetic hydrogels that consist of polyethylene-glycol (PEG) linkers of different molecular masses which are permanently cross-linked by hyperbranched polyglycerol (hPG) hubs~\cite{Herrmann2018}. Such synthetic hydrogels can be regarded as simple models for mucus, since they display size-dependent particle permeabilities~\cite{Herrmann2019, Witten2017}, similar to mucus. As diffusing particles we employ fluorescently labelled dextran molecules of varying sizes.
When using confocal laser-scanning fluorescence microscopy to investigate particle penetration into hydrogels, the sample can be oriented such that the hydrogel-bulk interface is either parallel~\cite{Marczynski2018} or perpendicular~\cite{Furter2019} to the optical axis, which makes no significant difference from a scanning perspective. However, for laterally extended samples like cell cultures that grow on a substrate, the parallel alignment causes the light path to span substantially larger distances, making this setup more prone to distortions in the imaging process. A perpendicular alignment, as employed in this work and sketched in Figure~\ref{fig:introduction}, is therefore preferable for biological samples~\cite{Furter2019} and is also compatible with future extensions of such penetration assays to mucus-producing cell cultures.

We investigate the filtering function of hydrogels by theoretical analysis of time-resolved concentration profiles of the labelled dextran molecules as they penetrate into the hydrogel. The employed numerical method allows for simultaneous extraction of free-energy and diffusivity profiles from relative concentration profiles at different times and is a significant extension of earlier methods~\cite{Schulz2017, Schulz2019, Lohan2020}, as it does not require absolute concentration profiles but works with relative, i.e. arbitrarily normalized, concentrations.
This is a crucial advantage, as often fluorescence intensity profiles are subject to significant perturbation due to e.g. laser light intensity fluctuations or fluorescence dye bleaching over the course of the experiment, and makes the often difficult conversion of measured intensity data into absolute particle concentrations obsolete.
Our method for the extraction of free-energy and diffusivity profiles from relative concentration profiles can be used for a wide range of different setups and systems. As a check on the robustness of the method, the extracted dextran bulk diffusivities are shown to agree well with fluorescence-correlation spectroscopy (FCS) data that are obtained separately. The obtained particle free energies and diffusivities inside the hydrogel are shown to obey scaling laws as a function of the dextran mass. The dextran free energy inside the hydrogel is described by a free-volume model based on repulsive steric interactions between the dextran molecules and the hydrogel linkers, which includes dextran as well as hydrogel linker flexibility. This model constitutes a modified size-filtering mechanism for repulsive particle-hydrogel interactions, according to which particle penetration into hydrogel pores
is assisted by the elastic widening of pores and the elastic shrinking of dextran molecules, and matches the extracted particle free energies in the hydrogel quantitatively. The model furthermore suggests that the hydrogel mesh size distribution is rather broad and that particle penetration is dominated by the fraction of large pores in the hydrogel.\\

\begin{figure*}
\includegraphics[width=\textwidth]{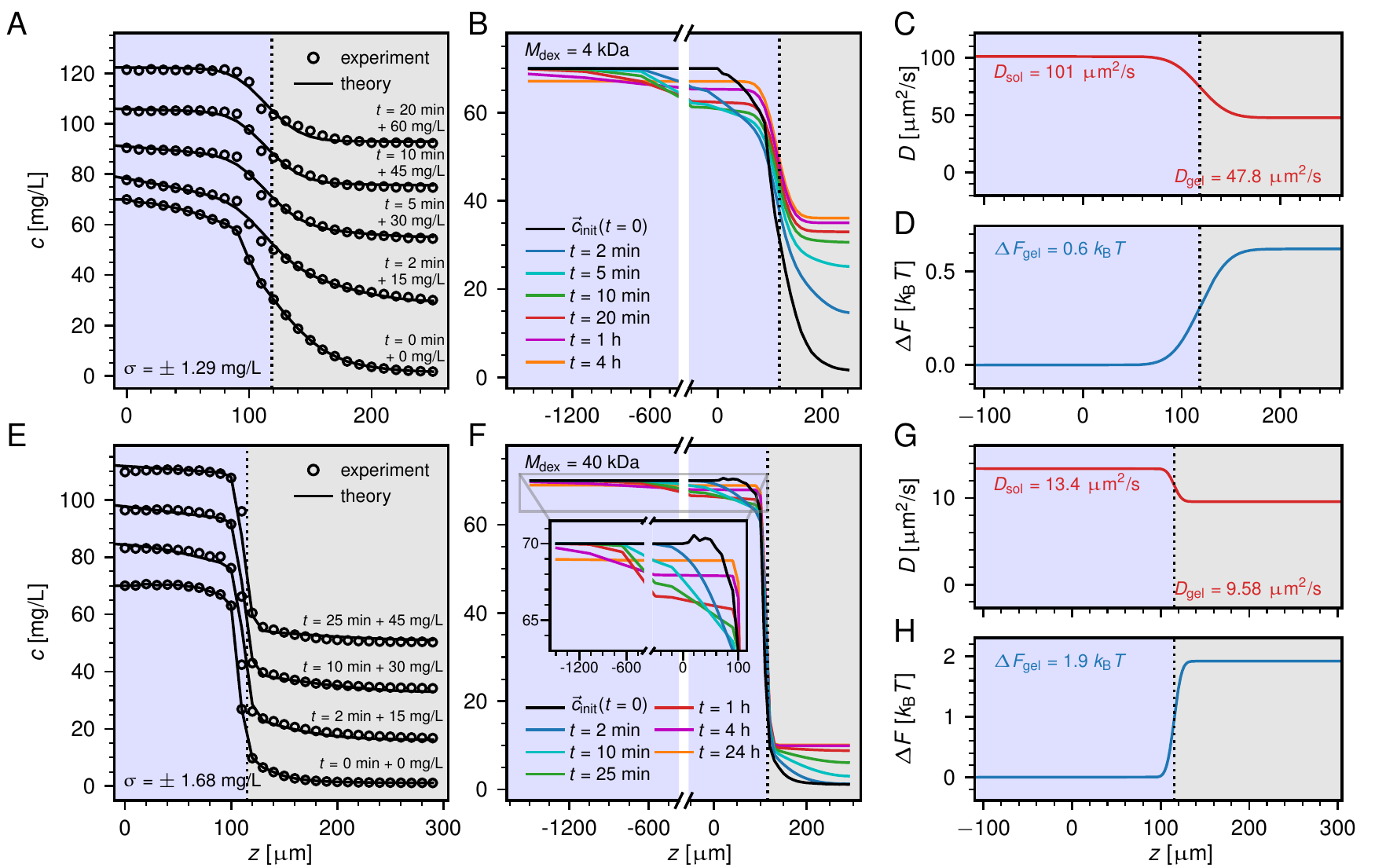}
\caption{Exemplary time-dependent dextran concentration profiles
from experimental measurements (circles) and numerical modeling (solid lines) for the \textit{hPG-G10} hydrogel. Results for the smallest dextran with  $M_{\text{dex}} = 4$~kDa in A-D are compared to results for $M_{\text{dex}} = 40$~kDa in E-H.
A\&E:~Experimental and modeled concentration profiles agree very accurately, note that concentration profiles are shifted vertically for better visibility.
B\&F:~Modeled concentration profiles are presented for a wide range of penetration times. The initial profile $\vec{c}_\text{init}$ (black line) is based on experimental data (see \hyperref[subsec:num_model]{Methods Section}~\ref{subsec:init_profile}).
C\&G:~Extracted diffusivity profiles, showing that the diffusivity in the hydrogel is only slightly reduced compared to the bulk solution.
D\&H:~Extracted free-energy profiles. Significant exclusion of dextran from the hydrogel is observed, with a stronger effect for the larger dextran.}
\label{fig:profiles}
\end{figure*}

\section{Results and Discussion}\label{sec:results}\noindent
Fluorescence intensity profiles of fluorescein isothiocyanate(FITC)-labeled dextran molecules penetrating into PEG-based hydrogels are analyzed using the procedure explained in the \hyperref[sec:methods]{Methods Section}. The analysis is based on numerical solutions of the one-dimensional generalized diffusion equation~\cite{risken2012fokker}

\begin{align*}
  \frac{\partial c(z, t)}{\partial t} =& \\
  \frac{\partial}{\partial z} &\left[ D(z)e^{-\beta F(z)} \frac{\partial}{\partial z}\left(c(z,t)e^{\beta F(z)} \right) \right]
  \numberthis
  \label{eq:FokkerPlanckEquation}
\end{align*}

\noindent
where $c(z, t)$ is the concentration at time $t$ and depth $z$ (see Figure~\ref{fig:introduction}), $D(z)$ and $F(z)$ are the spatially resolved diffusivity and free-energy profiles which the dextran molecules experience and $\beta=1/k_\text{B}T$ is the inverse thermal energy. While the diffusivity $D(z)$ describes the mobility of dextran molecules at position $z$, the free energy profile $F(z)$ uniquely determines the equilibrium partitioning of  dextran molecules.
The numerical solution of eq.~\eqref{eq:FokkerPlanckEquation} provides a complete model of the penetration process into the hydrogel and at the same time allows for extraction of the diffusivity and free energy profiles by comparison with experimentally measured concentration profiles.
A direct conversion of measured fluorescence intensities into absolute concentrations is often difficult due to drifts of various kinds. The method developed here circumvents this problem and allows for in-depth analysis of arbitrarily normalized concentration profiles, as explained in \hyperref[subsec:num_model]{Methods Section}~\ref{subsec:num_model}.
Complete profiles of free energies and diffusivities, both in the bulk and in the PEG hydrogel are obtained and the results for different hydrogels and dextran molecules of varying sizes will be analyzed in the following.\\

\textbf{Comparison Between Experimental and Modeled Concentration Profiles.}
Figure~\ref{fig:profiles}A\&E shows exemplary concentration profiles for
dextran molecules with molecular masses of $M_{\text{dex}} = 4$~kDa and $M_{\text{dex}} = 40$~kDa penetrating into the \textit{hPG-G10} hydrogel (see \hyperref[subsec:hydrogel]{Methods Section}~\ref{subsec:hydrogel}). Measurements are performed over a total time span of about 30 minutes and concentration profiles are recorded every 10 seconds, leading to a total of about 180 concentration profiles as input for the numerical extraction of the diffusivity and free-energy profiles.
The first measured concentration profile at $t=0$~min represents the start of the experiment, approximately 10~seconds after the dextran solution was applied onto the gel (see \hyperref[subsec:measurements]{Methods Section}~\ref{subsec:measurements}).
The numerically determined concentration profiles (lines) reproduce the experimental data (data points) very accurately, as seen in Figure~\ref{fig:profiles}A\&E. The deviation is estimated from the normalized sum of residuals, $\sigma$ (according to eq.~\eqref{eq:sigma}) which is below 2~mg/L for both measurements. A stationary concentration profile is obtained in the theoretical model only after 4 hours of penetration for the smaller 4~kDa dextran, see Figure~\ref{fig:profiles}B, for the larger dextran molecule the stationary profile is reached only after an entire day, see Figure~\ref{fig:profiles}F. These times significantly exceed the duration of the experiments.

The extracted diffusivity and free-energy profiles in Figure~\ref{fig:profiles}C, D, G, H, reveal the selective hydrogel permeability for dextran molecules of varying size. The free energy difference in the hydrogel is positive $\Delta F_\text{gel} > 0$ for both dextran sizes, indicating that dextran is repelled from the hydrogel. The dextran partition coefficient $K_\text{gel}$ between the hydrogel and the bulk solution is related to the change in the free energy $\Delta F_\text{gel}$ as

\begin{equation}
  K_\text{gel} = e^{-\beta\Delta F_\text{gel}}
  \label{eq:partition_coefficient}
\end{equation}

\noindent
According to eq.~\eqref{eq:partition_coefficient}, the obtained free energy differences $\Delta F_\text{gel} = 0.6~k_\text{B}T$ and $\Delta F_\text{gel} = 1.9~k_\text{B}T$, correspond to partition coefficients of about
$K_\text{gel}\approx 1/2$ and $K_\text{gel}\approx 1/7$ for the smaller and the larger dextran molecules, respectively, which illustrates a significant exclusion in particular for the larger dextran.
Compared to the partition coefficients, the diffusion constants in the hydrogel decrease only slightly as a function of the dextran mass. This suggests that the dextran molecules are only modestly hindered in their motion, a conclusion that will be rationalized by our elastic free-volume model further below.

\begin{figure}
\includegraphics[width=\columnwidth]{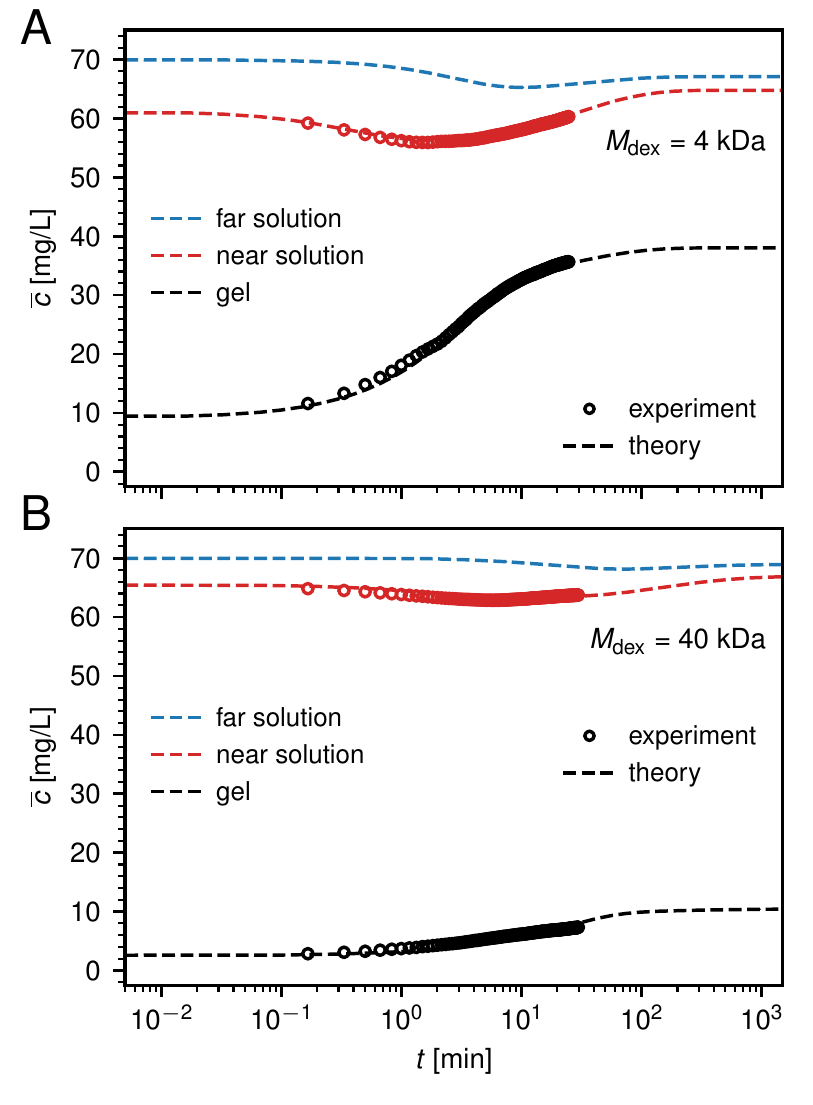}
\caption{Comparison of experimental results (circles) and modeling results based on the extracted diffusivity and free-energy profiles (lines) for the mean dextran concentration $\overline{c}$ over time in three different regions, the \textit{far solution} ($-z_{\text{top}} < z < 0$), the \textit{near solution} ($0 < z < z_{\text{int}}$) and the gel ($z_{\text{int}} < z < z_{\text{bot}}$), see Figure~\ref{fig:introduction}.
The systems are the same as shown in Figure~\ref{fig:profiles}. A non-monotonic dextran concentration is measured over time in the near and far solution regions.
The fact that $\overline{c}$ in the gel does not vanish for $t\rightarrow 0$
reflects that the first measurement at $t=0$ is done approximately 10~seconds after the application of the dextran solution onto the gel.
}
\label{fig:penetration_time}
\end{figure}

Figure~\ref{fig:penetration_time} shows the temporal evolution of the average dextran concentration $\overline{c}$ in three different regions, namely inside the gel for $z_{\text{int}} < z < z_{\text{bot}}$, in the \textit{near solution} for $0 < z < z_{\text{int}}$, and in the \textit{far solution} for $-z_{\text{top}} < z < 0$ for the same data shown in Figure~\ref{fig:profiles}.
The lines show the  predictions based on the extracted diffusivity and free-energy profiles, the circles the experimental data, which are not available in the \textit{far solution} range. The average concentration in the gel (black) increases monotonically and saturates after about one hour for both dextran sizes. Note that the stationary final concentration in the hydrogel is considerably less for the larger dextran with $M_{\text{dex}} = 40$~kDa. In contrast, the average concentration in the \textit{far solution} saturates more slowly and shows a slight non-monotonicity for both dextran masses (blue). This non-monotonicity is more pronounced in the \textit{near solution} (red) and is caused by the fact that dextran molecules diffuse quickly into the hydrogel from the \textit{near solution} in the beginning of the experiment, while the replenishment from the bulk solution takes a certain time, as also seen in the concentration profiles in Figure~\ref{fig:profiles}B\&F. Very good agreement between experiments and modeling results is observed.\\

% FIG4: Results
\begin{figure*}
\includegraphics[width=\textwidth]{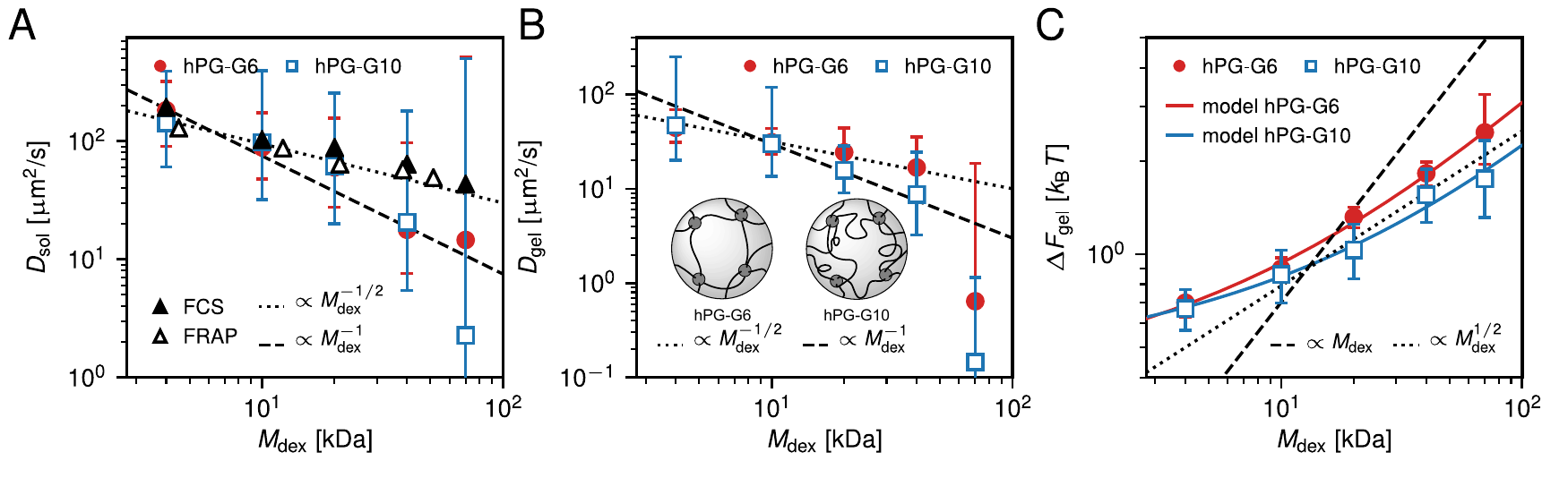}
\caption{Results for the diffusivity and free energy obtained from the experimental measurements as a function of dextran mass.
A:~Fitted diffusivities in the bulk solution (squares and circles) agree within the error with FCS data measured in the current work (solid black triangles) and with FRAP measurements from literature~\cite{Gribbon1998} (open black triangles).
B:~Fitted diffusivities in the hydrogel are reduced relative to the bulk values and are compared to different power laws.
C:~Dextran molecules are excluded from the hydrogel and $\Delta F_\text{gel} > 0$ for all dextran masses. For larger dextran molecules, $\Delta F_\text{gel}$ increases as a square root with the dextran mass. The results from the free-volume model of eq.~\eqref{eq:dF_full} (continuous lines) agree nicely with the measurements. Error bars have been estimated as explained in \hyperref[sec:suppinfo]{Section}~\ref{si-sec:errors} of the \hyperref[sec:suppinfo]{Supporting Information}. The inset in B presents a schematic depiction of the two different gels. Even though the \textit{hPG-G10} gel is composed of larger linkers, the mass density is larger than in the \textit{hPG-G6} gel, which results in an effectively smaller pore size.}
\label{fig:fit_results}
\end{figure*}

\textbf{Influence of Dextran Size on Hydrogel Penetration.}
The same analysis is performed for dextran molecules of molecular masses ranging from $M_\text{dex}=4~$kDa to $M_\text{dex}=70~$kDa that penetrate into PEG hydrogels with two different linker lengths, namely \textit{hPG-G6} with a PEG linker size of $M_\text{PEG} = 6$~kDa and \textit{hPG-G10} with $M_\text{PEG} = 10$~kDa. Figure~\ref{fig:fit_results} shows the extracted diffusivities and free energies, which result from averages over at least three experiments for each system, except for $M_\text{dex}=70~$kDa dextran, where only one experiment was performed.

Figure~\ref{fig:fit_results}A shows the bulk diffusivities $D_{\text{sol}}$ extracted from measured concentration profiles as colored symbols, in principle there should be no difference between results for \textit{hPG-G6} and \textit{hPG-G10}. A power law relation between the dextran mass and the diffusivity according to $D_\text{sol} \propto {M_\text{dex}}^{-\nu}$ is shown as straight lines for $\nu=1$ (broken line) and for $\nu=1/2$ (dotted line). An exponent of $\nu=1/2$ agrees nicely with our FCS data (solid black triangles; see \hyperref[subsec:FCS]{Methods Section}~\ref{subsec:FCS}) as well as with literature FRAP measurements~\cite{Gribbon1998} (open black triangles).
The value $\nu=1/2$ follows from combining the Stokes-Einstein relation $D_\text{sol} = k_\text{B}T/6\pi\eta_\text{w}r_0$ with the scaling of the dextran hydrodynamic radius according to
$r_0\propto{M_\text{dex}}^\nu$~\cite{Cheng2002, Bu1994} by assuming that the bulk solution is a theta solvent for dextran polymers~\cite{rubinstein2003polymer, Netz2003} (see \hyperref[sec:suppinfo]{Section}~\ref{si-sec:dex_mass_scaling} in the \hyperref[sec:suppinfo]{Supporting Information} for details).
The exponent $\nu=1/2$ is only expected for linear polymers, while dextran is in fact a branched polymer. The good agreement of FCS and FRAP data with the power law for $\nu=1/2$ suggests that the degree of branching is low~\cite{Granath1958} or that branching effectively compensates self-avoidance effects.
The dextran hydrodynamic radii estimated from the FCS measurements compare well with the values reported by the supplier, see Table~\ref{tab:dextran_size}.
The data for $D_\text{sol}$ obtained from the time-dependent dextran concentration profiles show rather large uncertainties, which is due to the fact that the concentration profiles are rather insensitive to the bulk diffusivities; they are within error bars consistent with our FCS results but do not allow extraction of the power-law scaling with any reasonable confidence.

\begin{table}
  \centering
  \begin{tabular}{| c | c | c |}
    \hline
    $M_\text{dex}$ & $r_0$ & $r_\text{FCS}$ \\
    \hline
    ~4~kDa & 1.4~nm & 1.5~nm \\
    \hline
    10~kDa & 2.3~nm & 2.7~nm \\
    \hline
    20~kDa & 3.3~nm & 3.2~nm\\
    \hline
    40~kDa & 4.5~nm & 4.3~nm \\
    \hline
    70~kDa & 6.0~nm & 6.4~nm \\
    \hline
  \end{tabular}
  \caption{Dextran hydrodynamic radius $r_0$ as reported by the supplier, in comparison to estimated hydrodynamic radius $r_\text{FCS}$ based on our FCS measurements using the Stokes-Einstein relation and the viscosity of water as $\eta_\text{w}=0.8\cdot 10^{-3}~$Pas.}
  \label{tab:dextran_size}
\end{table}

Values for the diffusion constant in the hydrogel $D_{\text{gel}}$ are compared to power laws with exponents $\nu=1/2$ and $\nu=1$ in Figure~\ref{fig:fit_results}B. The difference of the diffusion constants between the two different hydrogels is within the error bars, which
reflects the fact that the estimated mean hydrogel mesh-sizes, using a very simplistic hydrogel network model with a perfect cubic structure, are $l_0^\text{hPG-G6}=7.1~$nm and $l_0^\text{hPG-G10}=7.5~$nm (see \hyperref[subsec:mesh_size]{Methods Section}~\ref{subsec:mesh_size}) and thus quite similar to each other.
It is to be noted that for $M_\text{dex}\leq20$~kDa, the estimated mesh sizes are larger than twice the dextran hydrodynamic radii from Table~\ref{tab:dextran_size}, which would not suggest any dramatic confinement effect on the diffusion constant~\cite{Hansing2018Macromolecules}.
Interestingly, for the data where $M_\text{dex}\gtrsim 20$~kDa, the hydrogel with the larger linker length (\textit{hPG-G10}), which has a slightly higher mesh size, is seen to reduce the diffusion constant slightly more, which at first sight is counterintuitive.
This finding can be rationalized by the fact that the \textit{hPG-G10} gel has a higher mass density compared to the \textit{hPG-G6} gel (see \hyperref[subsec:hydrogel]{Methods Section}~\ref{subsec:hydrogel}), and thus the effective pore size is in fact substantially smaller. This is schematically illustrated in the inset in Figure~\ref{fig:fit_results}B.
A diffusivity scaling with an exponent $\nu=1$, which describes the data for \textit{hPG-G10} slightly better, could be rationalized by screened hydrodynamic interactions or by reptation-like diffusion~\cite{DeGennes1971}. In fact, a cross-over in the scaling of the diffusivity with increasing  hydrogel density from $\nu=1/2$ to $\nu=1$ has been described before for dextran penetrating into hydroxypropylcellulose~\cite{Bu1994}. However, because of the large error bars, extraction of the diffusivity scaling with respect to dextran mass in the two gels is not uniquely possible. This is mostly due to the fact that the diffusivities change rather mildly with varying dextran mass. This is why we do not attempt to model the scaling of the extracted  diffusivities, as was done elsewhere before~\cite{Hansing2016, Hansing2018a, Axpe2019}, but rather focus on the mechanism behind the
extracted free energy differences in the following.

\begin{figure*}
\includegraphics[width=\textwidth]{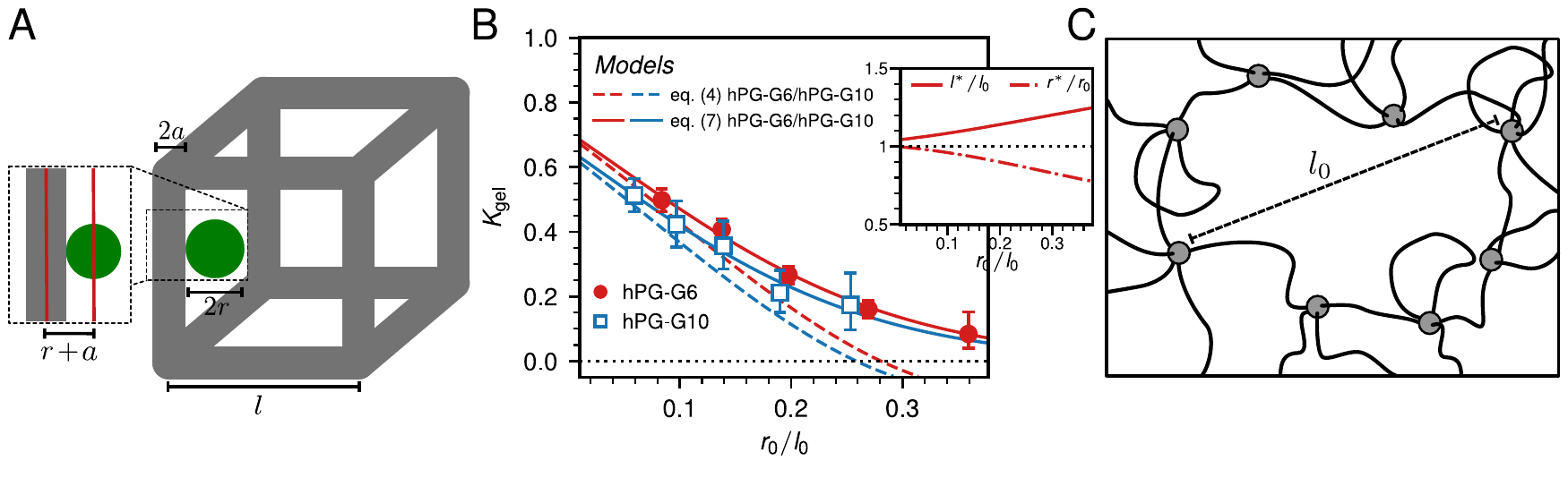}
\caption{Elastic free-volume model for the partitioning of a particle in a hydrogel.
A:~Schematic sketch of the cubic unit-cell model for the hydrogel, made up of connected linkers of length $l$ and a finite radius of $a$. The diffusing particle is modeled as a sphere of radius $r$. Both the particle and the linkers are elastic and can stretch or contract.
B:~Partition coefficient $K_\text{gel}$ extracted from the experimentally measured dextran concentration profiles (symbols) in comparison to the elastic free-volume model predictions according to eq.~\eqref{eq:dF_full} (solid lines). The results of the non-elastic model according to eq.~\eqref{eq:dF_vol} are shown as dashed lines. The inset shows the equilibrium values of $l^*$ and $r^*$ obtained for the \textit{hPG-G6} gel.
C: Illustration of a disordered pore in the hydrogel which has a mesh size $l_0$ and consists of more than four linkers (see also Fig.~\ref{fig:cubic_pores}).}
\label{fig:dF_model}
\end{figure*}

Figure~\ref{fig:fit_results}C shows the extracted values of $\Delta F_\text{gel}$ for the two hydrogels as a function of the dextran mass.
In all measurements we find $\Delta F_\text{gel} > 0$, which suggests exclusion of the dextran molecules from the hydrogel. Also the value of $\Delta F_\text{gel}$ increases with the dextran mass. Since dextran as well as the PEG-hPG based hydrogels are uncharged~\cite{Wang2008}, this exclusion must be due to steric repulsion, possibly enhanced by hydration repulsion
\cite{MatejPNAS2015,ManuPNAS2012}.\\

\textbf{Elastic Free-Volume Model for Dextran Penetration in Hydrogels.}
For the larger dextran molecules, the hydrogel with the smaller PEG linkers, \textit{hPG-G6}, displays a slightly stronger exclusion. The power law relation between the hydrogel free energy and dextran mass according to $\Delta F_\text{gel}\propto {M_\text{dex}}^\alpha$ with an exponent of $\alpha=1/2$
describes the data well for larger dextran masses $M_\text{dex}\gtrsim 20$~kDa, as shown by the dotted black line in Figure~\ref{fig:fit_results}C. This power law behavior is in fact compatible with a simplistic elastic free-volume model for the penetration of dextran molecules into hydrogels, which yields the solid lines and will be derived in the following.

The model geometry is sketched in Figure~\ref{fig:dF_model}A and consists
of a single dextran molecule of radius $r$ (green sphere) inside a cubic unit cell of the PEG based hydrogel (grey cylinders), similar to previous coarse-grained hydrogel models~\cite{Hansing2016, Hansing2018a, Hansing2018b}. The presence of the hPG hubs connecting the PEG linkers is neglected in the following. The dextran experiences a reduction of its free volume compared to the bulk solution, due to steric interactions with the PEG linkers. In the simple model geometry, the PEG linkers are located at the edges of the cubic unit cell and are modeled as impenetrable cylinders of radius $a$ and length $l$. The excluded volume $V_\text{ex}$ for dextran in the cubic unit cell consists of a quarter of each of the twelve cylinders at the edges. The accessible or free volume in the hydrogel $V_\text{free}$ depends on the sum of sphere radius $r$ and cylinder radius $a$ and is given by

\begin{align*}
  V_\text{free} &= V_\text{unit} - V_\text{ex}\\
               &= l^3 - \frac{12}{4}\pi(r+a)^2l + 2V_\text{cyl}
  \numberthis
  \label{eq:excluded_volume}
\end{align*}

\noindent
Here, $V_\text{unit}=l^3$ is the volume of the unit cell and $V_\text{cyl}=\frac{16}{3}(r+a)^3$ is the volume of two intersecting cylinders~\cite{Moore1974}, which is subtracted from the excluded volume to avoid over counting of the unit-cell corners.
The entropic contribution to the total free energy is given by

\begin{align*}
  &{\Delta F}_\text{vol} = -k_\text{B}T~
  \text{ln}\left(\frac{V_\text{free}}{V_\text{unit}}\right)
  \numberthis\label{eq:dF_vol}\\
  &= -k_\text{B}T~\text{ln}
  \left(1-3\pi{\left[\frac{r+a}{l}\right]}^2 +
  \frac{32}{3}{\left[\frac{r+a}{l}\right]}^3\right)
\end{align*}

\noindent
Since dextran and the PEG linkers are elastic polymers, they are both flexible and can deform. For small deformations, the polymers behave like Gaussian chains~\cite{rubinstein2003polymer, Netz2003}. The elastic deformation free energy for a cubic unit cell consisting of 12 equally deformed PEG linkers can be written as (for a detailed derivation see \hyperref[sec:suppinfo]{Section}~\ref{si-sec:peg_stretch} in the \hyperref[sec:suppinfo]{Supporting Information})

\begin{equation}
  {\Delta F}_\text{PEG} = \frac{12}{2}k_\text{B}T\left(\left[\frac{l}{l_0}\right]^2+\frac{1-4\left[\frac{l}{l_0}\right]^2}{2+\left[\frac{l}{l_0}\right]^2}\right)
  \label{eq:dF_PEG}
\end{equation}

\noindent
Here $l/l_0$ is the relative stretching of the PEG linkers, where $l_0$ denotes the edge length of the unit cell in the absence of dextran molecules.
The elastic deformation energy of dextran is obtained in the same fashion and reads

\begin{equation}
  {\Delta F}_\text{dex} = \frac{3}{2}k_\text{B}T\left(\left[\frac{r}{r_0}\right]^2+\left[\frac{r_0}{r}\right]^2-2\right)
  \label{eq:dF_dex}
\end{equation}

\noindent
where $r$ denotes the deformed dextran radius and the equilibrium dextran radius is denoted by $r_0$ and is taken from Table~\ref{tab:dextran_size}. The complete free energy follows as

\begin{align*}
  {\Delta F}_\text{gel}(r,l) &=\\
  {\Delta F}_\text{vol}&(r,l) + {\Delta F}_\text{PEG}(l) + {\Delta F}_\text{dex}(r)
  \numberthis
  \label{eq:dF_full}
\end{align*}

\noindent
The equilibrium free energy is given by the minimal value of this free energy expression, obtained for the optimal stretched unit cell length $l^*$ and the optimal dextran radius $r^*$, which are determined numerically.
The values of the unit cell length $l_0$ and the PEG linker thickness $a$ are adjusted by fits to the experimental data. The model results are shown in Figure~\ref{fig:dF_model}B in terms of the partition coefficient as solid lines and compared to the experiments (circles and squares) as a function of the length ratio $r_0/l_0$.
The inset shows the obtained equilibrium values for $l^*$ and $r^*$ for the \textit{hPG-G6} gel. A considerable stretching of PEG-linkers and compression of dextran is observed, which shows that elasticity effects of both PEG linkers and dextran molecules are important.

The fit to the experimental data yields $l_0^\text{hPG-G6}=16.7~$nm, $l_0^\text{hPG-G10}=23.7~$nm, $a_\text{hPG-G6}=3.4~$nm and $a_\text{hPG-G10}=5.4~$nm. The fit values of $a$ certainly represent an effective PEG linker radius and include the layer of tightly bound hydration water.
They are indeed, close to the respective equilibrium PEG radii $R_\text{PEG}=b_\text{fl}N_\text{PEG}^{3/5}/\sqrt{3}$, given as $R_\text{PEG}^\text{hPG-G6}=4.4~$nm and $R_\text{PEG}^\text{hPG-G10}=5.99~$nm, where $b_\text{fl}=0.4~$nm denotes the Flory monomer length~\cite{Liese2016} and $N_\text{PEG}$ is the respective number of PEG monomers.
In fact, the free-volume model yields estimates of the number of hydration waters per PEG monomer that scatter around 8, in rough agreement with literature values (see Figure~\ref{si-fig:peg_hydration} and \hyperref[sec:suppinfo]{Section}~\ref{si-sec:PEG_hyd} in the \hyperref[sec:suppinfo]{Supporting Information}).

The fit values for the unit cell length $l_0$ are significantly larger than the mean mesh size estimated based on eq.~\eqref{eq:mesh_estimate},
which for a perfectly ordered cubic lattice predicts $l_0^\text{hPG-G6}=7.1~$nm and $l_0^\text{hPG-G10}=7.5~$nm, but still considerably shorter than the PEG contour length $L=b_0^\text{PEG}N_\text{PEG}$, which is $L_\text{hPG-G6}=48.5~$nm and $L_\text{hPG-G10}=80.9~$nm, where $b_0^\text{PEG}=0.356~$nm is the PEG monomer length~\cite{Liese2016}. While the large unit cell lengths  obtained from the fit to the elastic free-volume model could reflect a substantial stretching of individual PEG polymers, there is no a priori reason why the linkers should be stretched to such a considerable fraction of their contour length. We therefore rationalize this surprising result in terms of a broad distribution of pore sizes that exhibit different topologies. To illustrate this, a random pore is schematically shown in Figure~\ref{fig:dF_model}C.
Based on the 3:1 number ratio of linkers and cross linkers in the hydrogel formulation (cf. \hyperref[subsec:hydrogel]{Methods Section}~\ref{subsec:hydrogel}), a perfectly cubic lattice could form, where each hub is connected to 6 different linkers. Such an ideal cubic connectivity is of course entropically highly unfavorable and the connectivity distribution of hubs, i.e. the distribution of the number of linkers that connect to one hub, will be rather broad and the network topology will be disordered, in which case the PEG end-to-end distance $R_\text{PEG}$ will be significantly smaller than the pore size $l_0$ (cf. also \hyperref[subsec:hydrogel]{Methods Section}~\ref{subsec:mesh_size}). While in a cubic lattice each cubic facet consists of four hubs and four linkers, the pores present in the actual hydrogel will show a broad distribution of the number of participating linkers. For illustration, the pore shown in Figure~\ref{fig:dF_model}C consists of eight linkers. Clearly, dextran molecules will tend to be located in larger pores in order to maximize their free volume, and therefore the fit parameters of our model will be dominated by the tail of the pore size distribution, which explains the large fit values for $l_0$.
This finding also allows to rationalize the larger extracted free energy barriers in the case of the \textit{hPG-G6} gel, even though the \textit{hPG-G10} gel mass density is higher (cf. Figure~\ref{fig:fit_results}C): The tail of the pore size distribution of the \textit{hPG-G10} gel presumably contains larger pores which can even stretch further to minimize the unfavorable dextran-PEG interactions. Clearly, the precise topology and compositional distribution of pores cannot be predicted by our analysis, our results should thus be merely interpreted as an indication of the presence of large pores and a disordered network topology.

An approximate non-elastic version of the free-volume model is obtained by neglecting the polymer deformation term and just keeping the excluded volume term, eq.~\eqref{eq:dF_vol}, which becomes accurate in the limit of $l_0\gg r_0$, where $r^*\approx r_0$ and $l^*\approx l_0$. These approximate results are shown as broken lines in Figure~\ref{fig:dF_model}B and describe the experimental data only for small values of $r_0/l_0$. When additionally approximating the logarithm in eq.~\eqref{eq:dF_vol}, the obtained expression for the free energy is similar to results derived for a random-fiber network~\cite{Giddings1968}.\\

\begin{figure}
\includegraphics[width=\columnwidth]{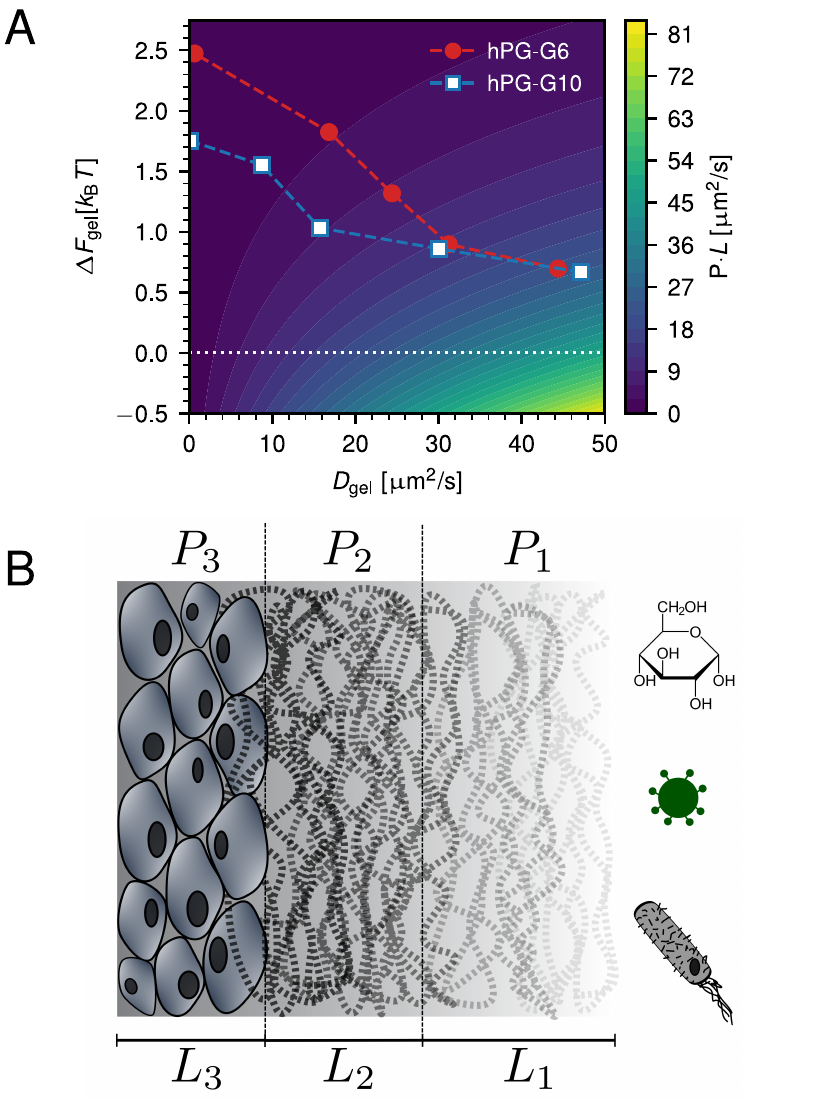}
\caption{A:~Normalized permeability coefficient $PL$ through a step-like single
hydrogel barrier of width $L$ as a function of the hydrogel free energy $\Delta F_\text{gel}$ and the hydrogel diffusivity $D_\text{gel}$ from eq.~\eqref{eq:permeability_approx}. High permeability is observed for low free-energy barriers and high diffusivities in the hydrogel. The symbols denote the experimental data from Figure~\ref{fig:fit_results}. Due to opposing trends in the free-energy barrier and the diffusivity, both hydrogels display comparable permeability coefficients.
B:~Schematic layered structure of a mucous membrane, as found in the stomach. Examples for different diffusors are shown, including nutrients such as glucose and pathogens such as virions or bacteria. The diffusors have to penetrate different layers of varying permeabilities to enter the tissue below the mucous membranes, the total permeability of a layered structure follows from eq.~\eqref{eq:layered_permeability}.}
\label{fig:permeabilities}
\end{figure}

\textbf{Derivation of Particle Permeabilities through Hydrogel Barriers.}
Permeation through biological barriers is quantified by the permeability coefficient $P$, which is defined as~\cite{Diamond1974}

\begin{equation}
  P(z_1, z_2) = \frac{J}{c(z_1)-c(z_2)}
  \label{eq:permeability_general}
\end{equation}

\noindent
where $c(z_1)$ and $c(z_2)$ are the particle concentrations at the two sides $z_1$ and $z_1$ of the barrier, and $J$ denotes the particle flux through the barrier. Based on the diffusion eq.~\eqref{eq:FokkerPlanckEquation}, the permeability can be written as (for a detailed derivation see \hyperref[sec:suppinfo]{Section}~\ref{si-sec:permeability} of the \hyperref[sec:suppinfo]{Supporting Information})

\begin{equation}
  \frac{1}{P(z_1, z_2)} = \int_{z_1}^{z_2} \frac{e^{\beta F(z)}}{D(z)}~dz
  \label{eq:permeability}
\end{equation}

\noindent
For a step-like barrier one obtains

\begin{equation}
  \frac{1}{P} = \frac{e^{\beta\Delta F_\text{gel}}}{D_\text{gel}}L
  \label{eq:permeability_approx}
\end{equation}

\noindent
Here $\Delta F_\text{gel}$ and $D_\text{gel}$ are the particle free energy relative to the solution and the diffusivity inside the hydrogel and $L$ denotes the width of the hydrogel barrier.

Figure~\ref{fig:permeabilities}A shows normalized permeability coefficients $P L$ for a single step-like barrier according to eq.~\eqref{eq:permeability_approx}, which are independent of the thickness of the barrier $L$, as a function of the gel free energy and the gel diffusivity.
The values extracted from the experimental data for different dextran molecules in the two gels from Figure~\ref{fig:fit_results} are indicated by data points.
Obviously, the highest permeability is observed for a low free-energy barrier and a high particle diffusivity, as is the case for the smallest dextran molecules (lower right corner in Figure~\ref{fig:permeabilities}A). On the other hand, permeation is hindered by either a high free-energy barrier or a low diffusivity in the hydrogel, both of which are observed for dextran molecules with larger molecular weights. Due to counterbalancing effects of stronger exclusion from the \textit{hPG-G6} gel and increased immobilization in the case of \textit{hPG-G10}, both hydrogels display comparable permeability coefficients for the chosen dextran molecular masses.\\

\section{Conclusion}\noindent
The method introduced in this paper allows for the simultaneous extraction of diffusivity and free-energy profiles of particles that permeate into spatially  inhomogeneous hydrogel systems, we demonstrate the method using concentration profile measurements of fluorescently labeled dextran molecules permeating into PEG-hPG-based hydrogels.
The advantage over alternative methods is that both diffusivity and free-energy profiles are obtained from a single experimental setup. This is important, as only the combination of diffusivity and free-energy profiles completely determines the diffusion of particles.

The extracted diffusivities and free-energies are analyzed in terms of scaling laws as a function of dextran mass and a modified elastic free-volume model is developed. This modified free-volume model includes the elasticity of PEG linkers and of the diffusing dextran molecules and quantitatively accounts for the extracted free energy values, demonstrating that elastic deformations of both the diffusor and the hydrogel network are important, in line with previous computational~\cite{Zhou2009, Godec2014, Kumar2019} and experimental studies~\cite{Mohamed2009}.
Our elastic free-volume model suggests that the hydrogel is topologically disordered and that dextran molecules preferentially move within exceptionally  large hydrogel pores, which are locally even more enlarged due to PEG strand elasticity.

Diffusional barriers in biological systems often show a layered structure, as previously demonstrated for skin~\cite{Schulz2017, Schulz2019, Lohan2020} and is also the case for mucous membranes, as found for instance in the gastrointestinal tract, schematically indicated in Figure~\ref{fig:permeabilities}B. For a layered system, eq.~\eqref{eq:permeability} shows that the individual piecewise constant permeability coefficients $P_i$ add up inversely as

\begin{align*}
  \frac{1}{P_\text{tot}} &= \sum_i\frac{1}{P_i} \\
  &= \sum_i\frac{e^{\beta\Delta F_i}}{D_i}L_i = \sum_i\frac{L_i}{D_iK_i}
  \numberthis
  \label{eq:layered_permeability}
\end{align*}

\noindent
where the sum goes over all layers, represented by their respective diffusion constants $D_i$, free energy values $\Delta F_i$ or partition coefficients $K_i$ and thicknesses $L_i$. Here, $P_\text{tot}$ denotes the total permeability, which is dominated by the smallest permeability in the inverse sum.

Figure~\ref{fig:permeabilities}B schematically illustrates permeation through a layered system which represents the mammalian stomach~\cite{Johansson2013}. The outermost layer of mucus is only loosely bound and characterized by the permeability $P_1$, it is followed by a layer of more tightly bound mucus, characterized by $P_2$, and adheres onto the first layer of epithelial cells, characterized by $P_3$.
The total thickness of this diffusional barrier is about a millimeter, with the two mucus layers spanning a few hundred micrometers only~\cite{Jordan1998}. Measurements in rat gastrointestinal mucosa suggest typical values of $L_1=109~\mu$m, $L_2=80~\mu$m and $L_3\approx L_2$~\cite{Atuma2001}, which are close to the range of gel thicknesses studied in this work.

The total permeability is determined by the free energies and the mobilities inside all layers. Nutrients for instance can easily penetrate through the epithelia of the gastrointestinal tract, displaying large permeabilities in the different layers. Pathogens on the other hand are in healthy environments kept from reaching the epithelium, due to low permeability in the tightly bound mucus layer ($P_2\ll P_1$)~\cite{Johansson2013}. From eq.~\eqref{eq:layered_permeability}, it is apparent that the lowest permeability in such a layered system dominates the total permeability, leading to an effective barrier function that for different particles can be caused
 by different parts of the layered barrier structure.

The method introduced in this work can be used to determine free-energy and diffusivity profiles, and thereby to predict effective permeabilities, of different kinds of fluorescently labeled molecules, particles or even organisms that penetrate into various layered systems, including systems that contain hydrogels and mucus. This will help to shed light on the underlying mechanisms of the function of biological barriers including mucous membranes.\\

\section{Methods}\label{sec:methods}
\begin{table*}[t]
  \centering
  \begin{tabular}{| l | c | c | c | c | c | c | c | c | c | c |}
    \hline
    & $n_\text{PEG}$ & ${V^\text{sol}_\text{PEG}}^a$ & $n_\text{hPG}$ & ${V^\text{sol}_\text{hPG}}^b$ & $V_\text{H$_2$O}$
    & $V^\text{sol}_\text{gel}$ & $V_\text{app}$ & $m_\text{app}$ & $n_\text{app}$ \\
    \hline
    \textit{hPG-G6} & 142~nmol & 10$~\mu$L & 47~nmol & 2.8~$\mu$L & 13.0~$\mu$L & 25.8~$\mu$L & 1~$\mu$L & 38~$\mu$g & 7.3~nmol \\
    \hline
    \textit{hPG-G10} & 84~ nmol & 10$~\mu$L & 28~nmol & 1.7~$\mu$L & 12.7~$\mu$L & 24.4~$\mu$L & 1~$\mu$L & 38~$\mu$g & 4.6~nmol \\
    \hline
  \end{tabular}
  \caption{Composition of the hydrogels used in this study. Here, $V^\text{sol}_\text{PEG}$ and $V^\text{sol}_\text{hPG}$ denote the volumes of the stock solutions, $V_\text{H$_2$O}$ is the volume of purified water added to the resulting gel solutions and $n_\text{PEG}$ and $n_\text{hPG}$ denote the amount of PEG linkers and hPG-hubs in the gel solutions.
  From the total resulting volume of the gel solutions $V_\text{gel}^\text{sol}$ only $V_\text{app}=1~\mu$L was placed as a gel spot on the glass substrate, leading to the applied amount $n_\text{app}$ and applied mass $m_\text{app}$.~$^a$Solution is of 8.5~wt\% for 6~kDa PEG and 8.4~wt\% for 10~kDa PEG.~$^b$hPG solution is of 5~wt\%.}
  \label{tab:gel_preparations}
\end{table*}

\subsection{Hydrogel Preparation.}\label{subsec:hydrogel}
The hydrogel is formed by cross-linking end-functionalized polyethylene glycol-bicyclo[6.1.0] non-4-yne (PEG-BCN) linkers with hyperbranched polyglycerol azide (hPG-N$_3$) hubs via strain-promoted azide-alkyne cycloaddition (SPAAC). The two macro-monomers PEG-BCN and hPG-N$_3$ are synthesized as previously described~\cite{Herrmann2018, Dey2016}. The \textit{click} reaction of binding the PEG-BCN linkers to the hPG-N$_3$ hubs works in water, at room temperature, without the addition of a catalyst or external activation like heat or UV radiation and without the formation of byproducts.
Two different sizes of PEG-BCN linkers are employed, having a molecular weight of either $M_\text{PEG}=6$ or $M_\text{PEG}=10$~kDa (for details about the mass distributions see \hyperref[sec:suppinfo]{Section}~\ref{si-sec:mass_distribution} of the \hyperref[sec:suppinfo]{Supporting Information}), the hydrogels are denoted as \textit{hPG-G6} and \textit{hPG-G10}, respectively.
The number ratio of the PEG-BCN linkers to the hPG-N$_3$ hubs (M$_\text{hPG}$ = 3~kDa, 20\% azide) is kept constant at 3:1 for both \textit{hPG-G6} and \textit{hPG-G10}. This ratio can ideally lead to a cubic lattice structure if each hPG-hub exactly binds to six PEG linkers. The chemical structure of the hPG-N$_3$ hubs, however, allows on average for eight binding sites, making the hydrogel presumably quite disordered.

The two components of the hydrogel are stored as aqueous stock solutions at concentrations of 8.5~wt\% (6~kDa PEG-BCN), 8.4~wt\% (10~kDa PEG-BCN) and 5~wt\% (hPG-N$_3$). To initiate hydrogel formation, they are mixed according to Table~\ref{tab:gel_preparations}. The resulting gel solution is thoroughly vortexed before being placed as 1~$\mu$L drops on the glass substrate. Both hydrogel solutions are adjusted to have the same mass concentration.
However, after drying and re-swelling on the glass substrate, volumes of the formed hydrogels are different and measured as $V^\text{hPG-G6}_\text{tot}=0.42~\pm~0.03~\mu$L and $V^\text{hPG-G10}_\text{tot}=0.31~\pm~0.04~\mu$L for \textit{hPG-G6} and \textit{hPG-G10}, respectively (see Figure~\ref{si-fig:gel_volumes} in \hyperref[sec:suppinfo]{Section}~\ref{si-sec:gel_volume} of the \hyperref[sec:suppinfo]{Supporting Information}).
This results in a final hydrogel concentration of 9~wt\% ($\approx$ 90 mg/mL) for \textit{hPG-G6} and 12~wt\% ($\approx$ 120 mg/mL) for \textit{hPG-G10}.

\subsection{Estimate of Mean Hydrogel Mesh Size.}\label{subsec:mesh_size}
\begin{figure}
\includegraphics[width=\columnwidth]{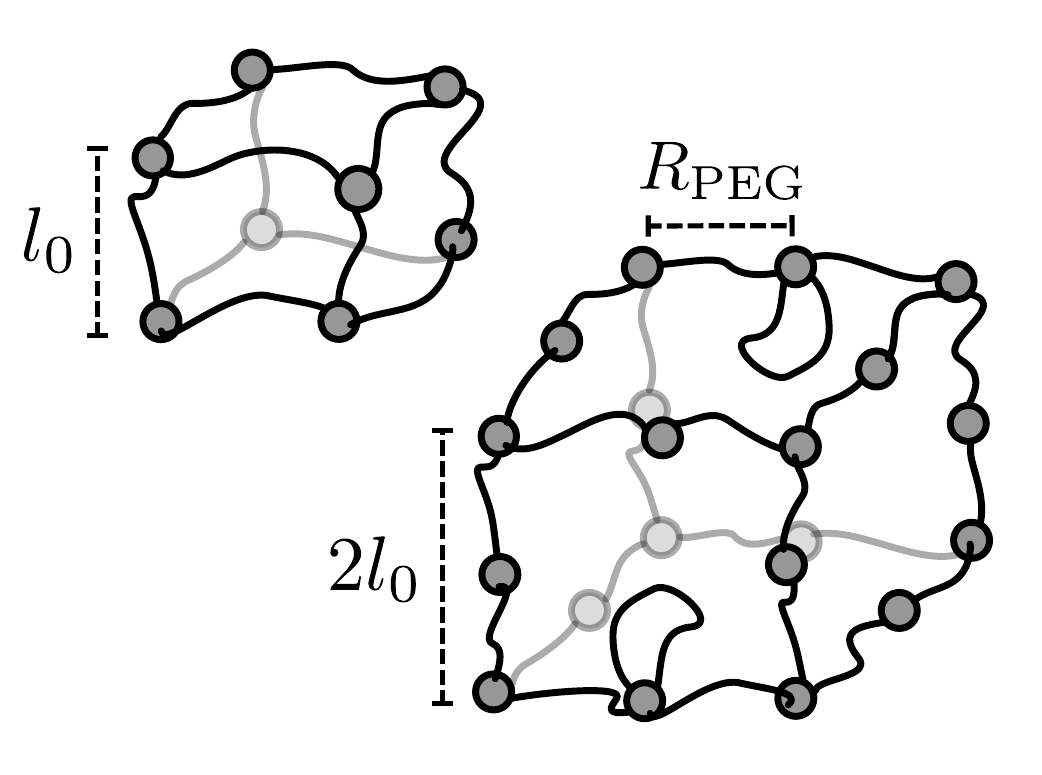}
\caption{A cubic pore with lower connectivity to the right, containing two instead of one PEG linker per edge, leads to an effectively larger unit cell length $l_0$ at the same PEG end-to-end distance $R_\text{PEG}$. Only for a perfect cubic lattice to the left, is the estimate of eq.~\eqref{eq:mesh_estimate} exact and $l_0=R_\text{PEG}$.}
\label{fig:cubic_pores}
\end{figure}

Assuming an idealized cubic hydrogel network structure, the mean mesh size can be easily estimated. The length of a cubic unit cell $l_0$ follows from the total gel volume $V_\text{tot}$ and the total number of hPG hubs $n_\text{hPG}^\text{tot}$ in mol as

\begin{equation}
  l_0 = \sqrt[\leftroot{-2}\uproot{2}3]{\frac{V_\text{tot}}{n_\text{hPG}^\text{tot}N_\text{A}}}
  \label{eq:mesh_estimate}
\end{equation}

\noindent
where $N_\text{A}$ is the Avogadro constant.
The total volumes for the re-hydrated gels are $V^\text{hPG-G6}_\text{tot} = 0.42~\mu$L and $V^\text{hPG-G10}_\text{tot} = 0.31~\mu$L as mentioned above.
The total number of hPG hubs is given as $n_\text{hPG}^\text{tot} = n_\text{hPG}\cdot V_\text{app}/V_\text{gel}^\text{sol}$, with the values from Table~\ref{tab:gel_preparations} for the respective gel and where we account for the fact that only $V_\text{app}=1~\mu$L of the total gel solution $V_\text{gel}^\text{sol}$ is applied onto the gel substrate.
This results in a rough estimates for the mesh size of $l_0^\text{hPG-G6}=7.1~$nm and $l_0^\text{hPG-G10}=7.5~$nm, which shows that even though PEG linkers of significantly different masses were used, the mesh sizes of the two gels differ only slightly.
In deriving eq.~\eqref{eq:mesh_estimate} one assumes an ideal hydrogel pore connectivity that corresponds to  a perfect cubic lattice. There is no reason why the hydrogel should consist of a perfect cubic lattice, on the contrary, entropy favors a disordered network topology. For cubic pores with lower connectivity, Fig.~\ref{fig:cubic_pores} illustrates how the pore size $l_0$ can increase for a fixed PEG end-to-end distance $R_\text{PEG}$. Thus, except for the case of an ideal cubic lattice, the estimated pore size $l_0$ will be larger than the estimate of eq.~\eqref{eq:mesh_estimate}, as indeed suggested by our elastic free-volume model.

\subsection{Dextran Preparation.}
Dextrans conjugated with the dye fluorescein isothiocyanate (FITC) are obtained from \textit{Sigma-Aldrich} as d4-FITC, d10-FITC, d20-FITC, d40-FITC and d70-FITC, the number stating the molecular weight in kDa of the commercial product. To remove unbound FITC from the dextran solutions, all batches are subjected to a desalting PD-10 column, which eliminates low-molecular weight compounds such as free FITC dye. This step is done according to the manufacturers recommendations and the column is equilibrated using phosphate buffer saline (PBS). Afterwards, the molecular weight distribution of all dextrans is determined by gel permeation chromatography (GPC) (see \hyperref[sec:suppinfo]{Section}~\ref{si-sec:mass_distribution} of the \hyperref[sec:suppinfo]{Supporting Information}).

\subsection{Penetration Assay of FITC-labeled Dextrans.}\label{subsec:measurements}
After preparation of the hydrogel solutions and purification of the dextrans (see above), penetration assays are performed with five different dextran solutions and two different gels.
For these assays, coverslips (Menzel \#1; VWR, Darmstadt, Germany) with a diameter of 25~mm and a thickness of 0.13-0.16~mm are thoroughly washed with water and absolute ethanol and subsequently dried under a stream of nitrogen. For every experiment, 1~$\mu$L of the respective hydrogel solution is placed in the center of the coverslip. The substrates with the applied gel spots are kept in a humid environment overnight, allowing hydrogel formation to be completed before the hydrogel spots are left to dry for 30~min at ambient conditions. Permeation experiments are performed within one day after hydrogel formation.
To start a permeation experiment, a home-made polydimethylsiloxane (PDMS) stamp (1 x 1 cm) prepared with a cylindrical cavity in the middle (5 mm diameter) is placed on the coverslip, so that the dried hydrogel is located in the middle of the stamp's cavity. The PDMS surrounding the dried hydrogel allows for the addition of solutions such as buffer or dextran. Prior to the measurement, 30~$\mu$L of PBS buffer are added to re-swell the hydrogel for 30~min, which typically creates hydrogel volumes of semi-spheroid shape with a base radius of 1050~$\mu$m and heights of about 150~$\mu$m for \textit{hPG-G10} and about 210~$\mu$m for \textit{hPG-G6} (see \hyperref[sec:suppinfo]{Section}~\ref{si-sec:gel_volume} of the \hyperref[sec:suppinfo]{Supporting Information}). Afterwards, the coverslip is mounted on a Leica SP8 confocal laser scanning microscope (CLSM; Leica, Wetzlar, Germany) and imaged using a 20x objective (0.75 HC PL APO water immersion objective with correction ring). In a first step, the hydrogel is visually identified by imaging the sample with a 488~nm laser and collecting the transmitted light using the transmission photomultiplier tube (PMT) of the CLSM, allowing to place the optical axis of the CLSM in the centre of the hydrogel and to place the focal plane 30~$\mu$m below the glass-hydrogel interface. After aligning the sample like this, the PBS buffer is removed from the cavity and replaced by 35~$\mu$L of the FITC-dextran solution (0.07~mg/mL for all dextrans).
This fixes the total length from the bottom of the glass dish at $z=z_\text{bot}$ to the air-water interface at $z=-z_\text{top}$, where $z=0$ corresponds to the end of the measurement region (see Figure~\ref{fig:introduction}A).
The total length of the solution is thus $z_{\text{tot}} = z_{\text{top}} + z_{\text{bot}} = 1780~\mu$m. The individual contributions to $z_{\text{tot}}$ vary, due to different gel thicknesses, changing the extent of the measured region, ranging from $z=0$ to $ z=z_\text{bot}$ (cf. also Figure~\ref{fig:introduction}A).

About 10 seconds after the application of the dextran solution, the spatial distribution of the FITC-based fluorescence intensity is measured using a z-stack that starts 30~$\mu$m below and ends 410~$\mu$m above the glass-hydrogel interface (with 10~$\mu$m increments). The recorded intensities are afterwards truncated to probe the spatial FITC distribution within the hydrogel starting from the glass bottom (located at $z_\text{bot}$) and extending about 100~$\mu$m into the bulk solution, away from the gel-water interface located at $z=z_\text{int}$ (cf. Figure~\ref{fig:introduction}A). In these measurements, the sample is excited at $\lambda$ = 488 nm and the emission is recorded between 500~nm and 550~nm using a PMT. For the $M_\text{dex}=4~$kDa to the $M_\text{dex}=40~$kDa dextrans, one z-stack is recorded every $\Delta t=10$~s, yielding time-resolved FITC distributions following the penetration of the dextran molecules into the hydrogel network over time.
For the $M_\text{dex}=70~$kDa dextrans a period of $\Delta t=30$~s is used instead, in order to account for the much smaller diffusion coefficient of the larger dextran molecules.
For all dextran types, measurements are performed at least three times with total measurement times of about 30 minutes, with the exception of the $M_\text{dex}=70~$kDa dextrans. Here only one measurement is performed for each gel but with a longer recording time of about 1 hour.

\subsection{Fluorescence Correlation Spectroscopy of FITC-labeled Dextrans.}\label{subsec:FCS}
Reference diffusion coefficients for the FITC-labeled dextran molecules in the bulk solution are obtained using fluorescence correlation spectroscopy (FCS). The measurements are performed on a Leica TCS SP5 II CLSM with a FCS set-up from \textit{PicoQuant}. The CLSM is equipped with an  HCX PL APO 63x/1.20 W CORR CS water immersion objective. Samples are put on high precision cover glasses (18 x 18~mm, 170 $\pm$ 5~$\mu$m thick) and excited with the 488 nm Argon laser line. The fluorescent light is passed through a 50/50 beam splitter with a lower wavelength cut-off of $\lambda=515$~nm. Both channels are detected separately with a single photon avalanche diode (SPAD). Afterwards a pseudo-cross correlation is performed between both channels to eliminate the influence of detector after-pulsing.
Prior to a measurement, the optical setup is calibrated with the water soluble Alexa-Fluor 488 dye. The correlated signal is fitted with two components and accounting for triplet states. The first component is fixed to a freely diffusing FITC-dye molecule where only the fraction is a fit parameter. The second component is set to a log-normal distributed species. The component fractions and means of distribution are fitted and the width of distribution is taken from previously performed gel permeation chromatography (GPC) measurements (for details about the fitting procedure see \hyperref[sec:suppinfo]{Section}~\ref{si-sec:FCS_fits} of the \hyperref[sec:suppinfo]{Supporting Information}). The fitted diffusion times are used to calculate the diffusion coefficients and hydrodynamic radii using the Stokes-Einstein relation.

\subsection{Numerical Model and Discretization.}\label{subsec:num_model}
Extending a previously introduced method~\cite{Schulz2017, Schulz2019, Lohan2020}, spatially resolved diffusivity and free energy profiles are estimated from experimentally measured concentration profiles. Numerical profiles are computed by discretizing the entire experimental setup from the glass bottom of the substrate to the air-water interface ($z_{\text{bot}}$ to $-z_{\text{top}}$ in Figure~\ref{fig:introduction}A).
In the regime where concentration profiles are measured ($z = 0$ to $z = z_{\text{bot}}$), the experimental resolution is used as the discretization width $\Delta z$ = 10~$\mu$m. For the range without experimental data ($z = 0$ to $z = -z_\text{top}$) in total six bins are employed. Two of those bins are spaced with  $\Delta z$ = 10~$\mu$m, for the other four bins, discretization spacings between $\Delta z$ =  300 - 400~$\mu$m are used, depending on the z-length measured in the respective experiment $z_{\text{bot}}$.
The z-dimension of the total system is the same for all experiments and given as $z_{\text{tot}} = z_{\text{top}} + z_{\text{bot}} = 1780~\mu$m. The experimentally measured region always extends from the glass bottom through the gel and at least 100$~\mu$m into the bulk solution, away from the hydrogel-bulk interface, which leads to values of $z_{\text{bot}}\approx300~\mu$m, depending on the exact thickness of the hydrogel in the respective measurement.

The numerical optimization problem is given by the cost function, which is defined as

\begin{align*}
\sigma^2(D, F, \vec{f}) :=& \\
\frac{1}{N\cdot M}\sum_{j=1}^{\text{N}} \sum_{i=1}^{\text{M}} &\left[ c^\text{num}_i(t_j) - f_{\text{j}}\cdot c^\text{exp}_i(t_j)\right]^2
\numberthis
\label{eq:sigma}
\end{align*}

\noindent
with $N$ the total number of experimental profiles, $M$ the total number of experimental data points per concentration profile and $\sigma^2(D, F, \vec{f})$ being the mean squared deviation between the experimental and numerical profiles. The diffusivity profile $D=D(z)$, the free energy landscape $F=F(z)$ and the vector containing all scaling factors (see below for details) $\vec{f} = (f_1, ..., f_j, ..., f_N)$ are all optimized to find the minimal value of $\sigma^2$. This non-linear regression is performed using the trust region method implemented in python's \textit{scipy} package~\cite{Branch1999}.

The numerical profiles \[\vec{c}_{\text{num}}(t_{\text{j}}) =  (c^\text{num}_1(t_{\text{j}}), ..., c^\text{num}_i(t_{\text{j}}), ..., c^\text{num}_M(t_{\text{j}}))^T\] are computed from the diffusivity and free energy profiles as

\begin{equation}
\vec{c}_{\text{num}}(t_{\text{j}}) = e^{Wt_{\text{j}}}\cdot \vec{c}_{\text{init}}
\label{eq:numerical_profiles}
\end{equation}

\noindent
where the rate matrix $W(D,F)$ is defined as

\begin{equation}
W_{i,k} = \frac{D_i+D_k}{2{\Delta z}^2}e^{-\frac{F_i-F_k}{2k_\text{B}T}}, \quad\text{with}\quad k=i\pm 1
\nonumber
\end{equation}

\noindent
as explained previously~\cite{Schulz2017}. Numerical profiles at time $t_j$ depend on
the initial profile $\vec{c}_{\text{init}}$ at $t = 0$, which is determined  as explained below.

The numerically computed profiles are fitted to the re-scaled experimental profiles $\vec{c}_{\text{exp}}(t_{\text{j}})$ at time $t_j>0$. The scaling factors $\vec{f}$ are obtained simultaneously from the fitting procedure and correct drifts in the experimentally measured fluorescence intensity profiles (see \hyperref[sec:suppinfo]{Section}~\ref{si-sec:drifts} of the \hyperref[sec:suppinfo]{Supporting Information}). As a check, the numerical model is compared to the analytical solution for a model with piece-wise constant values of the diffusivity and free energy in the respective regions. Results from the numerical model agree perfectly with those from the analytical solution (see \hyperref[sec:suppinfo]{Section}~\ref{si-sec:ana_sol} of the \hyperref[sec:suppinfo]{Supporting Information}).

\subsection{Construction of the Initial Concentration Profile.}\label{subsec:init_profile}
The initial profile $\vec{c}_{\text{init}}$, used for the computation of all later profiles according to eq.~\eqref{eq:numerical_profiles}, needs to cover the entire computational domain and is generated by extending the first experimentally measured profile $\vec{c}_{\text{exp}}(t = 0)$ into the bulk regime (from $z = 0$ to $z = -z_{\text{top}}$, cf. Figure~\ref{fig:introduction}A).
We define $t=0$ as the time of the first measurement, which is performed approximately 10~seconds after application of the dextran solution onto the gel-loaded substrate. For the extension, a constant initial concentration is assumed in the bulk, the value of which is taken as the experimentally measured value furthest into the bulk $c_0 := c^{\text{exp}}_1(t = 0)$ at $z=0$. This leads to the following expression used for the initial profile

\begin{align}
c^{\text{init}}_i :=
\begin{cases}
	c_0, & \text{if } -z_{\text{top}}\leq z_i \leq 0\\
	c^{\text{exp}}_i(t = 0), & \text{if } 0 < z_i \leq z_{\text{bot}}\\
\end{cases}
\label{eq:c0_profile}
\end{align}

\noindent
which by construction is continuous at $z=0$. The initial profiles used for the fit procedure are shown in Figure~\ref{fig:profiles}B and F as black lines. In order to obtain concentration profiles in physical units, we set the first measured value furthest into the bulk equal to the applied dextran concentration $c_0 = 0.07~$mg/mL.

\subsection{Free Energy and Diffusivity Profiles.}
The diffusivity $D(z)$ and free energy $F(z)$ profiles are assumed to change in a sigmoidal shape from their values in the bulk solution to their values in the hydrogel. This sigmoidal shape is modeled using the following expressions

\begin{align*}
D(z) &= \frac{D_{\text{sol}}+D_{\text{gel}}}{2} + \frac{D_{\text{sol}}-D_{\text{gel}}}{2} ~\text{erf}\left(\frac{z-z_{\text{int}}}{\sqrt{2}d_{\text{int}}}\right) \\
F(z) &= \frac{\Delta F_{\text{gel}}}{2} + \frac{\Delta F_{\text{gel}}}{2} ~\text{erf}\left(\frac{z-z_{\text{int}}}{\sqrt{2}d_{\text{int}}}\right)
\numberthis
\label{eq:sigmoid_DF}
\end{align*}

\noindent
where $\text{erf}(z) := 1/\sqrt{\pi} \int_{-z}^{z} e^{-z'^2}~dz'$ is the error function. The fit parameters $z_{\text{int}}$ and $d_{\text{int}}$ determine the transition position and width, respectively, and are the same for the free energy and diffusivity profiles. Since only free energy differences carry physical meaning, the free energy in the bulk solution is set to zero, so that $F_{\text{sol}}~=~0$.
The values of the diffusivity and free energy in the hydrogel and in the bulk solution are thus determined by fitting the five parameters of eqs.~\eqref{eq:sigmoid_DF}, namely $D_{\text{gel}}$, $\Delta F_{\text{gel}}$, $D_{\text{sol}}$, $z_{\text{int}}$ and $d_{\text{int}}$, to the experimentally measured concentration profiles.

Confidence intervals for the obtained parameters of $D_{\text{sol}}$, $D_{\text{gel}}$ and $\Delta F_{\text{gel}}$ are estimated by determining the parameter values that change $\sigma$ by not more than 50\% (for details see \hyperref[sec:suppinfo]{Section}~\ref{si-sec:errors} of the \hyperref[sec:suppinfo]{Supporting Information}). The error bars shown in Figure~\ref{fig:fit_results} are then obtained by averaging the confidence intervals over all measurements.

\begin{suppinfo}\label{sec:suppinfo}
\ref{si-sec:dex_mass_scaling}:~Scaling of Diffusion Constant with Dextran Size;
\ref{si-sec:peg_stretch}:~Expression for the Elastic Deformation Free Energy;
\ref{si-sec:PEG_hyd}:~Estimating PEG-Monomer Hydration Number;
\ref{si-sec:permeability}:~Permeability Coefficient;
\ref{si-sec:mass_distribution}:~Molecular Mass Distributions of PEG linkers and Dextran Molecules;
\ref{si-sec:gel_volume}:~Hydrogel Volume Reconstruction;
\ref{si-sec:FCS_fits}:~Fitting Procedure for FCS Measurements;
\ref{si-sec:drifts}:~Drifts in the Measured Fluorescence Intensity Data;
\ref{si-sec:ana_sol}:~Analytical Solution for Two-Segment System;
\ref{si-sec:errors}:~Error Estimate for Numerical Analysis.
\end{suppinfo}

\begin{acknowledgement}
The authors acknowledge funding by the Deutsche Forschungsgemeinschaft (DFG) via grant SFB 1449.
\end{acknowledgement}

% BIBLIOGRAPHY
\bibliography{main_refs}

\end{document}

\clearpage
\section{Scaling of Diffusion Constant with Dextran Size}\label{sec:dex_mass_scaling}
According to the Stokes-Einstein relation, the diffusion constant of dextran molecules in the bulk solution $D_\text{sol}$ is expected to scale with the dextran radius $r_0$ as

\begin{equation}
  D_\text{sol} \propto r_0^{-1}
\end{equation}

\noindent
Assuming the dextran polymer behaves like a freely jointed chain, its radius is related to the number of monomers $N_\text{dex}$ as~\cite{rubinstein2003polymer, Netz2003}

\begin{equation}
  r_0 = b_0^\text{dex}\sqrt{N_\text{dex}}
  \label{eq:freely_jointed_chain}
\end{equation}

\noindent
where $b_0^\text{dex}$ is the dextran monomer length and $N_\text{dex}$ can be estimated from the total molecular mass of a dextran molecule $M_\text{dex}$, when the monomer mass $M_\text{dex}^\text{mono}$ is known

\begin{equation}
  N_\text{dex} = \frac{M_\text{dex}}{M_\text{dex}^\text{mono}}
  \label{eq:mass_chain_length}
\end{equation}

\noindent
This leads to the following equality

\begin{equation}
  D_\text{sol} = \frac{k_\text{B}T}{6\pi\eta_\text{w} \frac{b_0^\text{dex}}{\sqrt{M_\text{dex}^\text{mono}}}\sqrt{M_\text{dex}}}
\end{equation}

\noindent
which gives rise to a scaling of $D_\text{sol}\propto M_\text{dex}^{-1/2}$.\\
\clearpage

\section{Expression for the Elastic Deformation Free Energy}\label{sec:peg_stretch}
The free energy cost for stretching a polymer chain from an initial equilibrium mean square end-to-end distance $\langle \vec{R}_0^2 \rangle$ to a larger end-to-end distance $\langle \vec{R}^2 \rangle$ can be written as~\cite{Netz2003, rubinstein2003polymer}

\begin{equation}
  \Delta F_\text{stretch} = \frac{3}{2}k_\text{B}T\frac{\langle \vec{R}^2 \rangle-\langle \vec{R}_0^2 \rangle}{\langle \vec{R}_0^2 \rangle}
  \label{eq:dF_stretch}
\end{equation}

\noindent
where $k_\text{B}T$ denotes the thermal energy. In the case of the PEG linkers we define the z-component of the end-to-end distance as $l := \langle R_z \rangle$, so that $l_0 := \langle R_{0,z} \rangle$. The PEG polymer chain is now only stretched in the z-direction, thus eq.~\eqref{eq:dF_stretch} reduces to

\begin{equation}
  \Delta F_\text{stretch} = \frac{3}{2}k_\text{B}T\frac{\langle R_x^2 \rangle + \langle R_y^2 \rangle + l^2 - \langle R_{0,x}^2  \rangle - \langle R_{0,y}^2  \rangle - l_0^2}{\langle \vec{R}_0^2 \rangle} = \frac{3}{2}k_\text{B}T\frac{l^2-l_0^2}{\langle \vec{R}_0^2 \rangle}
  \label{eq:dF_stretch_2}
\end{equation}

\noindent
since $\langle R_x^2 \rangle = \langle R_{0,x}^2 \rangle$ and $\langle R_y^2 \rangle = \langle R_{0,y}^2  \rangle$. As the PEG polymer chain performs a random walk in all three spatial dimensions, all components of the mean squared end-to-end distance contribute equally and so

\begin{equation}
  \frac{\langle \vec{R}_0^2 \rangle}{3} = \langle R_{0,x}^2  \rangle = \langle R_{0,y}^2  \rangle = \langle R_{0,z}^2  \rangle = l_0^2
  \label{eq:R_0_equal_components}
\end{equation}

\noindent
Together with eq.~\eqref{eq:dF_stretch_2}, eq.~\eqref{eq:R_0_equal_components} leads to the stretching free energy for a single PEG linker polymer chain

\begin{equation}
  \Delta F_\text{stretch} = \frac{1}{2}k_\text{B}T\left[\frac{l^2}{l_0^2}-1\right]
  \label{eq:dF_PEG_stretch}
\end{equation}

\noindent
For the compression of a polymer chain from an initially larger mean square end-to-end distance $\langle \vec{R}_0^2 \rangle$ to a smaller one $\langle \vec{R}^2 \rangle$ we write~\cite{Netz2003, rubinstein2003polymer}

\begin{equation}
  \Delta F_\text{compress} = \frac{3}{2}k_\text{B}T\frac{\langle \vec{R}_0^2 \rangle-\langle \vec{R}^2 \rangle}{\langle \vec{R}^2 \rangle}
  \label{eq:dF_compress}
\end{equation}

\noindent
In the same way as above, we only allow compression along the z-axis, which leads to

\begin{equation}
  \Delta F_\text{compress} = \frac{3}{2}k_\text{B}T\frac{\langle R_{0,x}^2  \rangle + \langle R_{0,y}^2 \rangle + l_0^2
  - \langle R_x^2 \rangle - \langle R_y^2 \rangle - l^2}{\langle R_x^2 \rangle + \langle R_y^2 \rangle + l^2} = \frac{3}{2}k_\text{B}T\frac{l_0^2-l^2}{\langle R_{0,x}^2  \rangle + \langle R_{0,y}^2 \rangle + l^2}
  \label{eq:dF_compress_2}
\end{equation}

\noindent
Using eq.~\eqref{eq:R_0_equal_components} to substitute the x- and y-components of the equilibrium end-to-end distance gives the expression for the compression free energy of a single PEG linker

\begin{equation}
  \Delta F_\text{compress} = \frac{1}{2}k_\text{B}T\frac{3l_0^2-3l^2}{2l_0^2 + l^2}
  \label{eq:dF_PEG_compress}
\end{equation}

\noindent
The total elastic deformation free energy per PEG linker is the sum of eq.~\eqref{eq:dF_PEG_stretch} and eq.~\eqref{eq:dF_PEG_compress}

\begin{equation}
    {\Delta F}_\text{PEG} = \frac{1}{2}k_\text{B}T\left(\left[\frac{l}{l_0}\right]^2+\frac{l_0^2-4l^2}{2l_0^2+l^2}\right)
  \label{eq:dF_PEG}
\end{equation}

\noindent
For the twelve PEG linkers of the hydrogel unit cell this leads to eq.~\eqref{main-eq:dF_PEG} of the main text.

\clearpage

\section{Estimating PEG-Monomer Hydration Number}\label{sec:PEG_hyd}
Based on eq.~\eqref{main-eq:partition_coefficient} and eq.~\eqref{main-eq:dF_full} from the main text, we obtain the following relation between the partition coefficient $K_\text{gel}$ and the volume accessible to the dextran diffusors $V_\text{free}$

\begin{equation}
  K_\text{gel} = \frac{V_\text{free}}{V_\text{unit}} e^{-\beta(\Delta F_\text{dex}+\Delta F_\text{PEG})}
  \label{eq:part_coeff_volume}
\end{equation}

\noindent
The volume inaccessible to the dextran molecules $V_\text{ex}$ (see eq.~\eqref{main-eq:excluded_volume} of the main text) is composed of a part occupied by the gel or dextran directly and a part due to tightly bound hydration water, so that

\begin{equation}
  V_\text{ex} = V_\text{unit} - V_\text{free} = V_\text{gel} + V_\text{hyd} + V_\text{dex}
  \label{eq:inacc_volume}
\end{equation}

\noindent
where $V_\text{gel}=V_\text{PEG}+V_\text{hPG}$ denotes the excluded volume due to both gel components, $V_\text{dex}=3\pi r^2l-\frac{32}{3}\pi r^3$ (with $r$ and $l$ as explained in Figure~\ref{main-fig:dF_model}A in the main text) denotes the excluded volume due to dextran, $V_\text{unit}=l^3$ is the unit cell volume and $V_\text{hyd}$ is the volume occupied by hydration water.
Since $r$ denotes the hydrodynamic radius of the spherical dextran, we assume $V_\text{hyd}$ to be the volume of hydration water only binding to the gel components.

The mass fraction $\Phi_\text{gel}$ of the gel components inside the hydrogel is defined as the ratio of the mass of the gel components $m_\text{gel}$ to the total mass $m_\text{tot}$, but since the mass density of the gel components is comparable to that of water, it also represents the fraction of inaccessible volume due to the gel components

\begin{equation}
  \Phi_\text{gel} := \frac{m_\text{gel}}{m_\text{tot}} \approx \frac{V_\text{gel}}{V_\text{unit}}
  \label{eq:vol_gel}
\end{equation}

\noindent
In the same fashion we can also estimate the fraction of inaccessible volume due to only the PEG linkers as

\begin{equation}
  \Phi_\text{PEG}=\frac{m_\text{PEG}}{m_\text{tot}}=\frac{n_\text{PEG}M_\text{PEG}}{m_\text{tot}}\frac{V_\text{app}}{V^\text{sol}_\text{gel}}\approx\frac{V_\text{PEG}}{V_\text{unit}}
  \label{eq:vol_PEG}
\end{equation}

\noindent
where we use the values for the number $n_\text{PEG}$ and the molar mass $M_\text{PEG}$ of the PEG linkers given in Table~\ref{main-tab:gel_preparations} of the Methods Section.
The factor of $V_\text{app}/V^\text{sol}_\text{gel}$ accounts for the fact that only $V_\text{app}=1$~$\mu$L of the total volume of prepared gel solution $V^\text{sol}_\text{gel}$ are actually placed on the gel spot for the experiments. The total mass of the hydrogel $m_\text{tot}$ is estimated from the measured hydrogel volumes by using the water mass density (cf. also \hyperref[sec:gel_volume]{Section}~\ref{sec:gel_volume}).
Combining eq.~\eqref{eq:part_coeff_volume}, eq.~\eqref{eq:inacc_volume}, and eq.~\eqref{eq:vol_gel} from above gives the following expression for volume fraction occupied by hydration water

\begin{equation}
  \frac{V_\text{hyd}}{V_\text{unit}} = 1-\Phi_\text{gel}-K_\text{gel} e^{\beta(\Delta F_\text{dex}+\Delta F_\text{PEG})}-3\pi\left(\frac{r}{l}\right)^2+\frac{32}{3}\left(\frac{r}{l}\right)^3
  \label{eq:vol_hyd}
\end{equation}

\noindent
The hydration water of eq.~\eqref{eq:vol_hyd} in principle binds to the entire hydrogel, meaning the hPG hubs and the PEG linkers. For a rough estimate, we neglect the presence of the hPG hubs and assume $V_\text{hyd}$ is the volume of water molecules hydrating only the PEG linkers. We can compute the fraction of hydration water per unit PEG volume from eq.~\eqref{eq:vol_PEG} and eq.~\eqref{eq:vol_hyd} as

\begin{equation}
  \frac{V_\text{hyd}}{V_\text{PEG}} = \frac{n_\text{hyd}v_\text{w}}{n_\text{PEG}v_\text{PEG}} = \frac{1-\Phi_\text{gel}-K_\text{gel} e^{\beta(\Delta F_\text{dex}+\Delta F_\text{PEG})}-
  3\pi\left(\frac{r}{l}\right)^2+\frac{32}{3}\left(\frac{r}{l}\right)^3}{\Phi_\text{PEG}}
  \label{eq:ratio_water_peg}
\end{equation}

\noindent
where $n_\text{hyd}$ is the number of hydration water molecules, $v_\text{w}$ is their partial volume, $n_\text{PEG}$ is the number of PEG linkers and $v_\text{PEG}$ is the PEG linker partial volume. The ratio between the partial volumes of water and the PEG linkers is approximated by the ratio of their molar masses as

\begin{equation}
  \frac{v_\text{w}}{v_\text{PEG}} \approx \frac{M_\text{w}}{M_\text{PEG}}
\end{equation}

\noindent
with the water molar mass $M_\text{w} = 18~$g/mol and the molar mass of the respective PEG linker $M_\text{PEG}$ (see Methods Section). From eq.~\eqref{eq:ratio_water_peg}, we can now compute the number of hydration waters per PEG linker molecule as

\begin{equation}
  \frac{n_\text{hyd}}{n_\text{PEG}} = \frac{1-\Phi_\text{gel}-K_\text{gel} e^{\beta(\Delta F_\text{dex}+\Delta F_\text{PEG})}-
  3\pi\left(\frac{r}{l}\right)^2+\frac{32}{3}\left(\frac{r}{l}\right)^3}{\Phi_\text{PEG}}\frac{M_\text{PEG}}{M_\text{w}}
  \label{eq:per_PEG_linker}
\end{equation}

\noindent
In order to obtain the number of hydration waters per PEG-monomer  we simply divide eq.~\eqref{eq:per_PEG_linker} by the respective number of PEG-monomers per linker $N_\text{PEG}$, which we obtain from the ratio of total linker mass $M_\text{PEG}$ and PEG-monomer mass $M_\text{PEG}^\text{mono}=44~$g/mol, as $N_\text{PEG}=M_\text{PEG}/M_\text{PEG}^\text{mono}$. The number of hydration waters per PEG-monomer can thus be obtained as

\begin{equation}
  \frac{n_\text{hyd}}{n_\text{PEG}^\text{mono}} = \frac{1-\Phi_\text{gel}-K_\text{gel} e^{\beta(\Delta F_\text{dex}+\Delta F_\text{PEG})}-
  3\pi\left(\frac{r}{l}\right)^2+\frac{32}{3}\left(\frac{r}{l}\right)^3}{\Phi_\text{PEG}}\frac{M_\text{PEG}^\text{mono}}{M_\text{w}}
  \label{eq:per_PEG_mono}
\end{equation}

\begin{figure*}
 \includegraphics[]{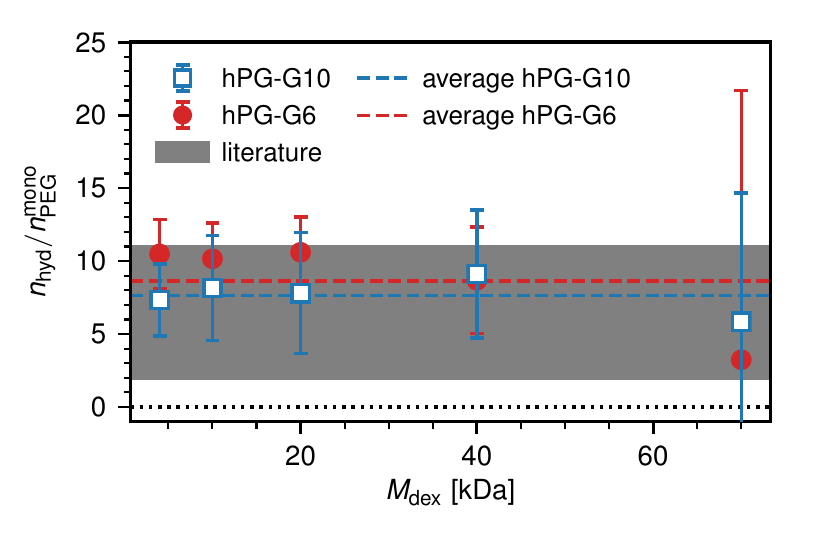}
 \caption{Estimated number of water molecules per PEG monomer $n_\text{hyd}/n_\text{PEG}^\text{mono}$ from the obtained values of $K_\text{gel}$ for the two hydrogels based on eq.~\eqref{eq:per_PEG_mono}.
 The estimated values scatter around $n_\text{hyd}/n_\text{PEG}^\text{mono}=8$ and agree with the range of values reported in the literature indicated as the grey shaded area, ranging from $n_\text{hyd}/n_\text{PEG}^\text{mono}=2$ to $n_\text{hyd}/n_\text{PEG}^\text{mono}=11$~\cite{Huang2001, Shikata2006, Kaatze1978, Bieze1994, Zwirbla2005}.}
 \label{fig:peg_hydration}
\end{figure*}

\noindent
With the values of $K_\text{gel} e^{\beta(\Delta F_\text{dex}+\Delta F_\text{PEG})}$,
$V_\text{dex}/V_\text{unit}=3\pi\left(\frac{r}{l}\right)^2-\frac{32}{3}\left(\frac{r}{l}\right)^3$, $\Phi_\text{gel}$ and $\Phi_\text{PEG}$ for the two hydrogels from the main text, eq.~\eqref{eq:per_PEG_mono} allows us to estimate the number of hydration waters per PEG monomer for all measurements.
Figure~\ref{fig:peg_hydration} shows the results of the calculation for each of the two hydrogels. Estimated values scatter around 8 water molecules per PEG monomer, with a slight dependence on the PEG linker length and dextran mass. Depending on the employed experimental method, values reported in the literature vary, ranging from 2 to 11 water molecules per PEG monomer~\cite{Huang2001, Shikata2006, Kaatze1978, Bieze1994, Zwirbla2005}.
Additionally, an increase of the hydration waters per monomer has been observed, as a function of the polymerization degree~\cite{Branca2002}. The values obtained from our estimates, all lie within the range of values reported in the literature, as  indicated by the grey shaded area in Figure~\ref{fig:peg_hydration}. This further corroborates our methodology and specifically the model for the free energy of eq.~\eqref{main-eq:dF_full}, since the estimate of eq.~\eqref{eq:per_PEG_mono} is based on this model.

\clearpage

\section{Permeability Coefficient}\label{sec:permeability}
The definition of the permeability coefficient $P$ is, as stated in the main text~\cite{Diamond1974}

\begin{equation}
  P(z_1, z_2) := \frac{J}{c(z_1)-c(z_2)}
  \label{eq:permeability}
\end{equation}

\noindent
with the stationary flux $J$ and the equilibrium concentrations on  both sides of the barrier $c(z_1)$ and  $c(z_2)$. From the generalized diffusion equation~\eqref{main-eq:FokkerPlanckEquation} in the main text, one obtains the stationary flux for the case of $\partial c(z,t)/\partial t=0$ as

\begin{equation}
  J = D(z)e^{-\beta F(z)}\frac{\partial}{\partial z}\left( c(z,t)e^{\beta F(z)}\right)
  \label{eq:eq_flux}
\end{equation}

\noindent
as a function of the diffusion constant $D(z)$ and the free energy landscape $F(z)$ across the barrier. After rearranging eq.~\eqref{eq:eq_flux} and integrating from one side of the barrier from $z_1$ to the other $z_2$, we obtain the following relation

\begin{equation}
  J\int_{z_1}^{z_2}\frac{e^{\beta F(z)}}{D(z)}~dz = c(z_1,t)e^{\beta F(z_1)}-c(z_2,t)e^{\beta F(z_2)}
\end{equation}

\noindent
We now assume that the free energy value is the same on both sides of the barrier and additionally set it to zero as reference so that $F(z_1)=F(z_2)=0$ and thus

\begin{equation}
  J\int_{z_1}^{z_2}\frac{e^{\beta F(z)}}{D(z)}~dz = c(z_1,t)-c(z_2,t)
\end{equation}

\noindent
which, in combination with eq.~\eqref{eq:permeability}, gives eq.~\eqref{main-eq:permeability} used in the main text

\begin{equation}
  \frac{1}{P} = \int_{z_1}^{z_2}\frac{e^{\beta F(z)}}{D(z)}~dz
\end{equation}

\section{Molecular Mass Distributions of PEG linkers and Dextran Molecules}\label{sec:mass_distribution}
The dextran molecular mass distribution is characterized using gel permeation chromatography (GPC), the results of which are presented in Table~\ref{tab:dex_mass_distribution}. GPC measurements are performed on an \textit{Agilent} device (1100er series) with a PSS Suprema column (pre-column, 1x with poresize of 30~\AA, 2x with poresize of 1000~\AA, all of them with a particle size of 10~$\mu$m), with pullulan as calibration standard, and ethylene glycol as internal standard. As solvent, H$_2$O with 0.1~M NaNO$_3$ is used. For the characterization of the PEG linker mass distribution matrix assisted laser desorption ionization (MALDI) on a \textit{Bruker Ultraflex} II is performed. The obtained values are shown in Table~\ref{tab:PEG_mass_distribution}.
From the number average molecular mass $M_\text{n}$ and the weight average molecular mass $M_\text{w}$ the polydispersity index $PDI$ is determined as

\begin{equation}
  PDI = \frac{M_\text{w}}{M_\text{n}}
  \label{eq:PDI}
\end{equation}

\noindent
The dextran molecules show a considerable dispersion regarding the molecular mass, while the PEG linker mass distribution is rather uniform.

\begin{table}[h]
  \centering
  \begin{tabular}{| c | c | c | c |}
    \hline
    $M_\text{dex}$ [kDa] & $M_\text{w}$ [kDa] & $M_\text{n}$ [kDa] & ~$PDI$~ \\
    \hline
    ~4 & 3.55 & 2.32 & 1.53 \\
    \hline
    10 & 9.55 & 5.55 & 1.72 \\
    \hline
    20 & 16.5 & 9.42 & 1.75 \\
    \hline
    40 & 35.7 & 19.5 & 1.84 \\
    \hline
    70 & 61.9 & 50.1 & 1.24 \\
    \hline
  \end{tabular}
  \caption{Results obtained from GPC measurements of the different dextran molecules.}
  \label{tab:dex_mass_distribution}
\end{table}

\begin{table}[h]
  \centering
  \begin{tabular}{| c | c | c | c |}
    \hline
    $M_\text{PEG}$ [kDa] & $M_\text{w}$ [kDa] & $M_\text{n}$ [kDa] & ~$PDI$~ \\
    \hline
    ~6 & 6.29 & 5.90 & 1.07 \\
    \hline
    10 & 11.3 & 11.3 & 1.00 \\
    \hline
  \end{tabular}
  \caption{Molecular weights and polydispersity of the PEG linkers as measured in MALDI experiments.}
  \label{tab:PEG_mass_distribution}
\end{table}
\clearpage

\section{Hydrogel Volume Reconstruction}\label{sec:gel_volume}
\begin{figure*}
\includegraphics[]{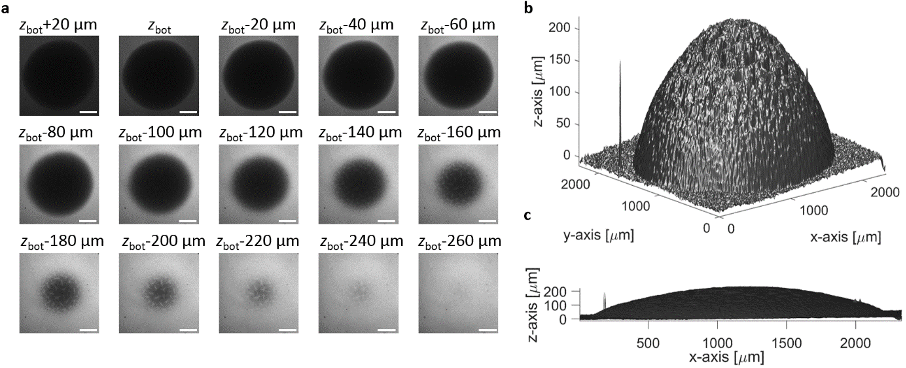}
\caption{In these experiments, the hydrogels have been equilibrated using PBS buffer (as described in Methods Section of the main text) followed by application of a $M_\text{dex}=70~$kDa dextran solution. The dextran is too large to penetrate into the hydrogel as evidenced by using confocal fluorescence microscopy to record multiple sample sections (at different z-heights as indicated in (a)). In these images, the hydrogel appears as a black circle, the radius of which decreases with increasing distance to the glass interface (located at the z-height $z_\text{bot}$). The bright areas correspond to FITC-dextran solution in the bulk. These images can be used to extract the position of the glass interface and hydrogel in space. A 3D representation and a side view of this hyperplane are given in (b) and (c), respectively, showing that the hydrogels possess a semi-elliptical shape with radii being on the order of 1050 $\mu$m (major axis) and 150 – 210 $\mu$m (minor axis). This figure shows a representative measurement for the \textit{hPG-G6} gel. The scale bars in (a) correspond to 500 $\mu$m.}
\label{fig:gel_volumes}
\end{figure*}

The hydrogel volume is determined using confocal laser scanning microscopy as follows. The hydrogels are first equilibrated using PBS buffer, followed by injection of the $M_\text{dex}=70~$kDa FITC-labeled dextran and recording of 3D dextran concentration profiles using confocal microscopy (covering an imaged volume of 2,304 x 2,304 x 0.5 mm$^3$). The permeation measurements (shown in the main text) reveal that this dextran is too large to penetrate the hydrogel, so that the hydrogel can be identified in these 3D concentration profiles based on an exclusion of FITC-labeled dextran (i.e., absence of FITC fluorescence; see Figure~\ref{fig:gel_volumes}).
In order to quantify the hydrogel volume, the (2D) hyperplane at which the FITC intensity has dropped to 50\% of its bulk value is determined using home-written scripts in \textit{Matlab} (\textit{MathWorks}, Natick, MA). This hyperplane indicates the positions, at which the point spread function of the confocal microscope is equally filled by the FITC-dextran bulk solution and either the hydrogel or the supporting substrate, and therefore allows for extraction of the exact locations of the substrate and hydrogel interface within the sample. Furthermore, as the substrate is a flat glass slide, the hydrogel-substrate interface is extracted from this data by first fitting a plane to the regions corresponding to the substrate-bulk interface (i.e., in regions far away from the hydrogel spot) and by interpolating the position of this plane underneath the hydrogel.
This procedure allows for extraction of the entire hydrogel boundary, allowing determination of the hydrogel volume by numerical integration. This yields volumes of $V^\text{hPG-G6}_\text{tot}=0.42~\pm~0.03~\mu$L and $V^\text{hPG-G10}_\text{tot}=0.31~\pm~0.04~\mu$L for the two hydrogels.

\clearpage

\section{Fitting Procedure for FCS Measurements}\label{sec:FCS_fits}
An accurate description of the dextran autocorrelation functions (ACFs) G($\tau$) required to use at least 2 components, which originate from the fluorescence emission of the FITC-labeled dextrans and, in addition, from residuals of free FITC molecules (i.e., not being conjugated to dextran). The FCS ACF G($\tau$) was therefore described using the equation

\begin{equation}
  G(\tau) = T(\tau)\cdot\left(G_\text{F}(\tau)+G_\text{DF}(\tau)\right)
\label{eq:FCS_1}
\end{equation}

\noindent
in which $G_\text{F}(\tau)$ and $G_\text{DF}(\tau)$ give the contributions from free FITC molecules and FITC-labeled dextrans, respectively, while the term $T(\tau)$ accounts for triplet state dynamics according to

\begin{equation}
  T(\tau) = \frac{\left(1-\Phi_\text{T}+\Phi_\text{T}e^{-\frac{\tau}{\tau_\text{T}}}\right)}{1-\Phi_\text{T}}
\label{eq:FCS_2}
\end{equation}

\noindent
with $\Phi_\text{T}$ denoting the fraction of molecules in the triplet state and $\tau_\text{T}$ the corresponding decay time of triplet states~\cite{Widengren1994}.

The contribution of the free FITC molecules was modeled using the theoretically derived ACF for free diffusion in 3D

\begin{equation}
  G_\text{F}(\tau) = \rho_\text{F}
  \left(1+\frac{\tau}{\tau_\text{F}}\right)^{-1}
  \left(1+\frac{\tau}{\tau_\text{F}\kappa^2}\right)^{-\frac{1}{2}}
\label{eq:FCS_3}
\end{equation}

\noindent
in which $\rho_\text{F}$ denotes the average number density of free FITC molecules in the confocal readout volume, $\tau$ the lag time of the ACF, $\kappa$ the ratio of axial $r_\text{z}$ to radial extension $r_\text{xy}$ of the confocal readout volume, and $\tau_\text{F}$ the decay time~\cite{rigler2012fluorescence}.
The value of $\tau_\text{F}$ was determined from calibration measurements on FITC molecules diffusing in buffer and $\kappa$ was fixed to 6~\cite{Ruttinger2008}, so that only the density $\rho_\text{F}$ was a free parameter when fitting the component $G_\text{F}(\tau)$ to experimentally determined ACF.

Since the dextrans showed a log-normal size distribution, as observed in the GPC measurements, the component of FITC-labeled dextrans of the ACF, $G_\text{DF}(\tau)$, was modeled by the superposition

\begin{equation}
  G_\text{DF}(\tau) = \sum_i p_\text{i} G_\text{DF,i}(\tau)
\label{eq:FCS_4}
\end{equation}

\noindent
using the log-normally distributed weights

\begin{equation}
  p_\text{i} := \frac{1}{\tau_\text{DF,i}\sigma_\text{DF}\sqrt{2\pi}}e^{-\frac{\left(\text{ln}(\tau_\text{DF,i})-\mu_\text{DF}\right)^2}{2\sigma_\text{DF}^2}}
\label{eq:FCS_5}
\end{equation}

\noindent
and the corresponding ACF contributions

\begin{equation}
  G_\text{DF,i}(\tau) := \rho_\text{DF}
  \left(1+\frac{\tau}{\tau_\text{DF,i}}\right)^{-1}
  \left(1+\frac{\tau}{\tau_\text{DF,i}\kappa^2}\right)^{-\frac{1}{2}}
  \label{eq:FCS_6}
\end{equation}

\noindent
The parameter $\sigma_\text{DF}$, which determines the broadness of the log-normal distribution, was determined by matching the FCS-related polydispersity index ($PDI$), defined by

\begin{equation}
  PDI_\text{FCS} = \frac{\sum_i p_i M_i^2}{\left(\sum_i p_i M_i\right)^2}
  \label{eq:FCS_7}
\end{equation}

\noindent
using

\begin{subequations}
\begin{align}
  M_i &\propto R_i^3\\
  R_i &= \frac{k_\text{B}T}{6\pi\eta D_i}\\
  D_i &= \frac{r_\text{xy}^2}{4\tau_\text{DF,i}}
  \label{eq:FCS_8}
\end{align}
\end{subequations}

\noindent
where the $PDI$ of the dextran mass distribution was determined using GPC. The parameters $\rho_\text{DF}$ and $\tau_\text{DF}=e^{\mu_\text{DF}}$ denote the average number density of FITC-labeled dextran molecules in the confocal readout volume and their average decay time, respectively, and were determined when fitting the component $G_\text{DF}(\tau)$ to the experimentally determined ACF.

Fitting the 2-component model therefore yields information about the number densities of free FITC molecules and FITC-labeled dextran molecules in the confocal readout volume ($\rho_\text{F}$ and $\rho_\text{DF}$, respectively) and their decay times $\tau_\text{F}$ and $\tau_\text{DF}$, which can be translated into diffusion coefficients using

\begin{subequations}
\begin{align}
  D_\text{F} &= \frac{r_\text{xy}^2}{4\tau_\text{F}}\\
  D_\text{DF} &= \frac{r_\text{xy}^2}{4\tau_\text{DF}}
  \label{eq:FCS_9}
\end{align}
\end{subequations}

\noindent
and into hydrodynamic radii using the Stokes-Einstein relation~\cite{rigler2012fluorescence}.
\clearpage

\section{Drifts in the Measured Fluorescence Intensity Data}\label{sec:drifts}
\begin{figure*}
\includegraphics[]{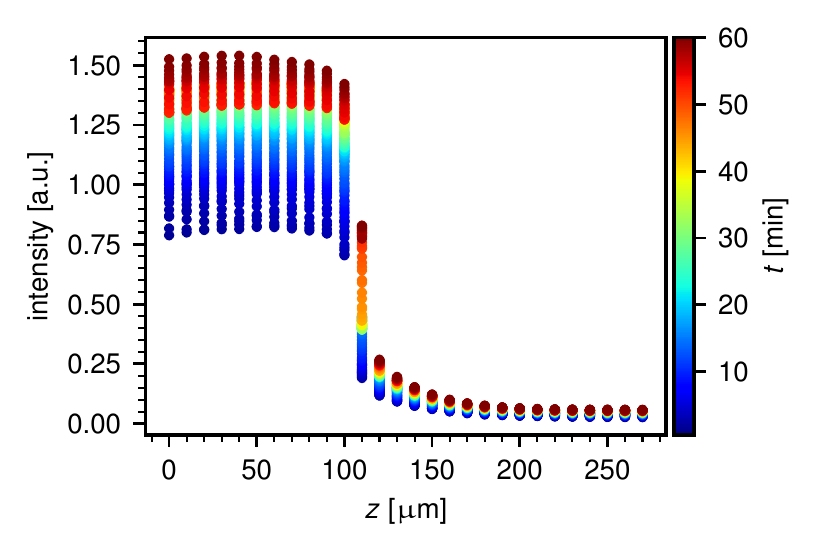}
\caption{Raw fluorescence intensity data from experiments of $M_\text{dex}=70~$kDa dextrans in combination with the \textit{hPG-G10} hydrogel. A significant change in the signal over time is observed in the probed part of the bulk solution, even though almost no penetration of the dextrans into the hydrogel is apparent. This drift in the experimentally measured signal is overcome by the numerically determined re-scaling factors.}
\label{fig:raw_profiles}
\end{figure*}

\begin{figure*}
\includegraphics[]{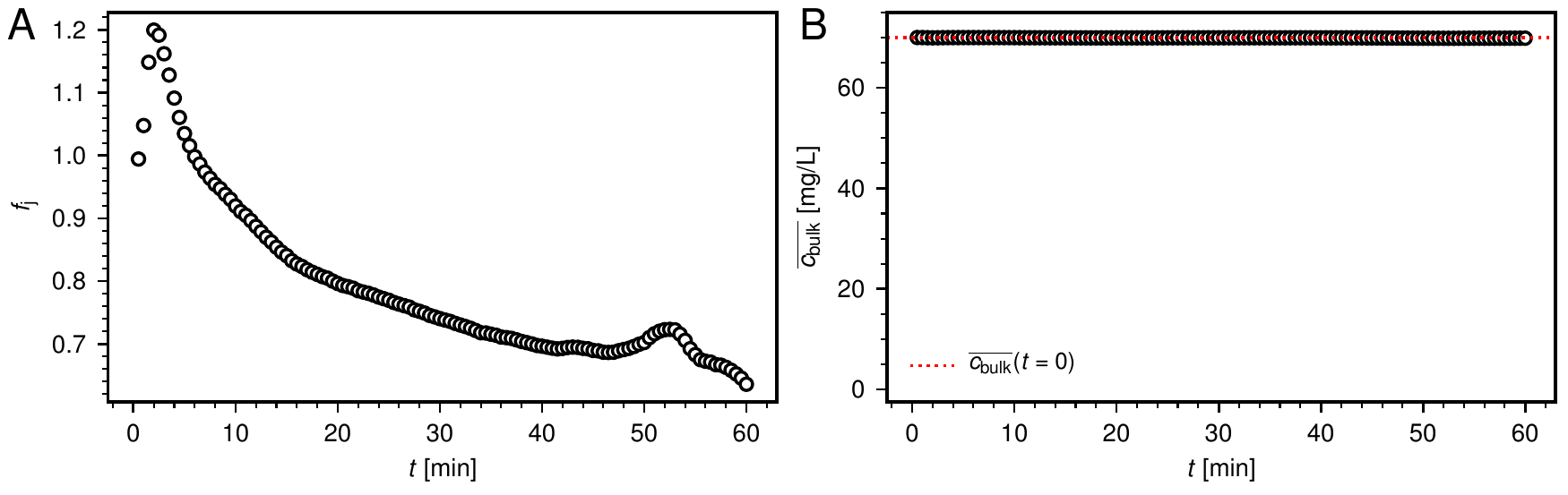}
\caption{A:~Set of re-scaling factors $\vec{f}$ for every measured concentration profile obtained from numerical analysis of experimental data from $M_\text{dex}=70~$kDa dextrans at the \textit{hPG-G10} hydrogel interface. Decreasing re-scaling factors counteract the drift observed in the raw experimental data. Additionally, peaks in the re-scaling factor distribution are observed, counteracting shorter fluctuations in the fluorescence intensity. B:~Average bulk concentration $\overline{c_\text{bulk}}$ computed according to eq.~\eqref{eq:avg_bulk_concentration} for the same measurements. The bulk concentration remains constant in this measurement, since almost no dextran penetrates into the hydrogel.}
\label{fig:scalings}
\end{figure*}

The experimentally measured fluorescence intensity data displays a continuous drift in the signal in all recorded measurements, which is likely due to an automatic re-adjustment of the laser intensity in the used setup. An example of the observed drift in the raw un-scaled signal is shown in Figure~\ref{fig:raw_profiles}, recorded for $M_\text{dex}=70$~kDa dextran molecules at the \textit{hPG-G10} interface. Even though almost no penetration of the large dextran molecules into the hydrogel is observed, the fluorescence intensity in the probed part of the bulk solution changes significantly over time. In order to obtain physical values for the dextran concentration, the measured profiles are being re-scaled during the fitting procedure. The obtained re-scaling factors for every measured concentration profile $\vec{f}$ decline over time, thus overcoming the constant increase of signal intensity due to the drift (see Figure~\ref{fig:scalings}A). Additionally, smaller changes in the fluorescence intensity are apparent. Since robust results are obtained by employing this re-scaling routine, this suggests that the entire information about the diffusion process is present in the relative shape of the concentration profiles.

The obtained re-scaling factor can additionally be used, to estimate the experimental bulk concentration $c_\text{bulk}$ far away from the hydrogel interface, based on the experimentally measured profiles alone, without using the numerically computed dextran distributions. The total amount of dextran in the system $C_{\text{tot}}$ is computed from the first concentration profile as $C_{\text{tot}} = \int_{-\infty}^{\infty} c(z, t=0)~dz$, where $c(z, t=0)$ was approximated by $c^\text{init}_i$ according to eq.~\eqref{main-eq:c0_profile} in the main text.
An average experimental concentration in the bulk region $\overline{c}_{\text{bulk}}(t_{\text{j}})$ can then be estimated from the fitted re-scaling factors as

\begin{equation}
\overline{c}_\text{bulk}(t_{\text{j}}) = \frac{C_{\text{tot}} - \sum_{i=1}^M f_{\text{j}}\cdot c^\text{exp}_i(t_j)\cdot\Delta z_i}{z_{\text{top}}}
\label{eq:avg_bulk_concentration}
\end{equation}

\noindent
Values for $\overline{c}_\text{bulk}(t_\text{j})$ are shown in Figure~\ref{fig:scalings}B and are virtually constant for the exemplary measurement of $M_\text{dex}=70$~kDa dextrans, as is expected due to the absence of penetration into the hydrogel.

\clearpage

\section{Analytical Solution for Two-Segment System}\label{sec:ana_sol}
\begin{figure*}[b]
\includegraphics[]{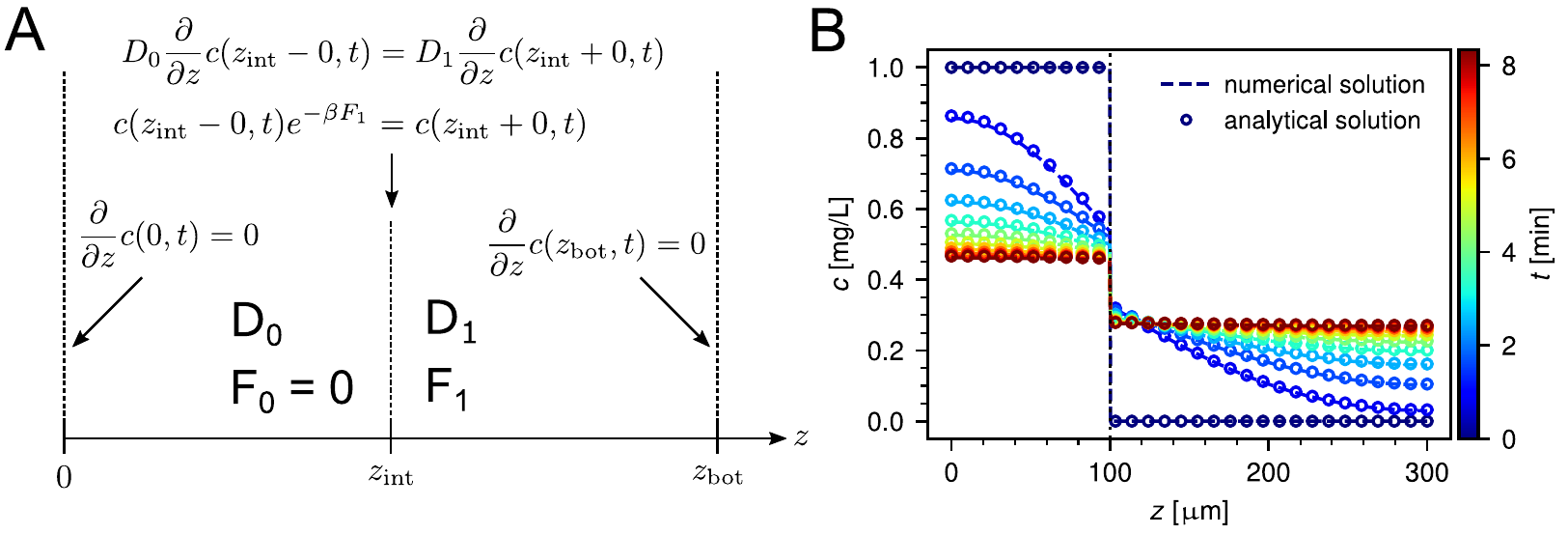}
\caption{Approximation of the dextran hydrogel setup as a two-segment system, which can be solved analytically. A:~Two-segment system with different diffusion constants and a jump in the free energy at the interface $z_\text{int}$. The used boundary conditions are also indicated. B:~Comparison of the numerical model and the analytical solution of the system explained in~A, with values for the parameters of $z_\text{int}=100~\mu$m, $z_\text{bot}=300~\mu$m, $c_0=1$, $D_0=50~\mu$m$^2/s$, $D_1=100~\mu$m$^2/s$ and $F_1=0.5~k_\text{B}T$.}
\label{fig:ana_comparison}
\end{figure*}

Simplifying the hydrogel-water setup as a two-box system with piece-wise constant values of the free energy and diffusion constant in the two regions allows for an analytical solution of the diffusion problem. The modeled system with the corresponding boundary conditions is sketched in Figure~\ref{fig:ana_comparison}A.

We solve the following diffusion equation in each of the two segments

\begin{equation}
  \frac{\partial}{\partial t}c(z,t) = D(z)\frac{\partial^2}{\partial z^2}c(z,t)
\label{eq:diffusion_eq}
\end{equation}

\noindent
where the diffusion constant $D(z)$ has a different value in each of the two regions

\begin{equation}
  D(z) =
  \begin{cases}
        D_0, & \text{if } 0\leq z\leq z_\text{int} \\
        D_1, & \text{if } z_\text{int}< z\leq z_\text{bot} \\
  \end{cases}
\end{equation}

\noindent
as does the free energy $F(z)$, which we set to zero in the left segment as reference

\begin{equation}
  F(z) =
  \begin{cases}
        F_0=0, & \text{if } 0\leq z\leq z_\text{int} \\
        F_1, & \text{if } z_\text{int}< z\leq z_\text{bot} \\
  \end{cases}
\end{equation}

\noindent
At the interface $z_\text{int}$, the flux needs to be continuous due to mass conservation, while the jump in the free energy leads to a jump in the concentration profile $c(z=z_\text{int},t)$. This defines the boundary conditions at $z_\text{int}$ as

\begin{subequations}
\begin{align}
  \lim_{z\nearrow z_\text{int}}D_0\frac{\partial}{\partial z}c(z,t) &=
  \lim_{z\searrow z_\text{int}}D_1\frac{\partial}{\partial z}c(z,t)\\
  \lim_{z\nearrow z_\text{int}}c(z,t)e^{-\beta F_1} &=
  \lim_{z\searrow z_\text{int}}c(z,t)
\end{align}
\label{eq:bc_interface}
\end{subequations}

\noindent
Since we are modeling a closed system, the edges at $z=0$ and $z=z_\text{bot}$ are reflecting boundaries with

\begin{subequations}
\begin{align}
  \frac{\partial}{\partial z}c(z=0,t) &= 0\\
  \frac{\partial}{\partial z}c(z=z_\text{bot},t) &= 0
\end{align}
\label{eq:bc_edges}
\end{subequations}

\noindent
Initially, the diffusors are only present in the left segment, modeling the bulk solution. This defines our initial condition as

\begin{equation}
  c(z, t=0) =
  \begin{cases}
        c_0, & \text{if } 0\leq z\leq z_\text{int} \\
        0, & \text{if } z_\text{int}< z\leq z_\text{bot} \\
  \end{cases}
\end{equation}

\noindent
We now solve eq.~\eqref{eq:diffusion_eq} by means of Laplace transformation. To this end, we use the single sided Laplace transform in time, defined as $\hat{f}(s) := \int_0^\infty f(t)e^{-st}dt$, where $s$ is the complex variable in Laplace space $s=\sigma + i\omega$. This converts the partial differential equation~\eqref{eq:diffusion_eq} into an ordinary differential equation of second order

\begin{equation}
  \left[s-D(z)\frac{\partial^2}{\partial z^2}\right]\hat{c}(z,s) = c(z,t=0)
\label{eq:diffusion_eq_LPT}
\end{equation}

\noindent
The general solution of eq.~\eqref{eq:diffusion_eq_LPT} for the two regions reads

\begin{equation}
  \hat{c}(z, s) =
  \begin{cases}
        a_1e^{\lambda_0z}+a_2e^{-\lambda_0z}+\hat{c}_\text{p}, & \text{if } 0\leq z\leq z_\text{int} \\
        a_3e^{\lambda_1z}+a_4e^{-\lambda_1z}, & \text{if } z_\text{int}< z\leq z_\text{bot}\\
  \end{cases}
  \label{eq:sol_LPT}
\end{equation}

\noindent
where we define $\lambda_\text{i} := \sqrt{\frac{s}{D_\text{i}}}, i=0,1$ and $\hat{c}_\text{p}:=\frac{c_0}{s}$.
The coefficients $a_i$ of eq.~\eqref{eq:sol_LPT} are determined by solving the system of linear equations obtained by Laplace transforming the boundary conditions of eq.~\eqref{eq:bc_interface} and eq.~\eqref{eq:bc_edges} and substituting the general solution (eq.~\eqref{eq:sol_LPT}). After some algebra, the solution to the posed problem is obtained as

\begin{equation}
  \hat{c}(z, s) =
  \begin{cases}
        \hat{c}_\text{p}
        \frac{
        K\cdot\text{tanh}(\lambda_1(z_\text{bot}-z_\text{int}))
        \left[
        \text{cosh}(\lambda_0z_\text{int})-\text{cosh}(\lambda_0z)
        \right] +
        \text{sinh}(\lambda_0z_\text{int})\sqrt{\delta}}{
        K\cdot\text{tanh}(\lambda_1(z_\text{bot}-z_\text{int}))
        \text{cosh}(\lambda_0z_\text{int})+
        \text{sinh}(\lambda_0z_\text{int})\sqrt{\delta}
        }
        , & \text{if } 0\leq z\leq z_\text{int} \\
        \hat{c}_\text{p}
        \frac{
        K\cdot\text{cosh}(\lambda_1(z_\text{bot}-z))
        \text{tanh}(\lambda_0z_\text{int})\sqrt{\delta}}{
        K\cdot\text{sinh}(\lambda_1(z_\text{bot}-z_\text{int}))+
        \text{tanh}(\lambda_0z_\text{int})
        \text{cosh}(\lambda_1(z_\text{bot}-z_\text{int}))\sqrt{\delta}
        }, & \text{if } z_\text{int}< z\leq z_\text{bot}\\
  \end{cases}
  \label{eq:full_laplace_transform}
\end{equation}

\noindent
where $\delta:=\frac{D_0}{D_1}$ and $K:=e^{-\beta F_1}$.\\\indent
The solution in Laplace space (eq.~\eqref{eq:full_laplace_transform}) is then transformed into real space by use of the Mellin integral

\begin{align*}
  c(z,t) &= \frac{1}{2\pi i}\int_{s=\sigma-i\infty}^{s=\sigma+i\infty} \hat{c}(z,s)e^{st}~ds\\
  &= \frac{e^{\sigma t}}{2\pi}\int_{-\infty}^{+\infty} \hat{c}(z,\sigma+i\omega)e^{i\omega t}~dw
  \numberthis
\end{align*}

\noindent
where the last integral was solved numerically through the inverse discrete Fourier transform.

Figure~\ref{fig:ana_comparison}B shows a comparison of the analytical solution and the numerical model for an exemplary parameter set of $z_\text{int}=100~\mu$m, $z_\text{bot}=300~\mu$m, $c_0=1$~mg/L, $D_0=50~\mu$m$^2/s$, $D_1=100~\mu$m$^2/s$ and $F_1=0.5~k_\text{B}T$,
mimicking a slight immobilization and repulsion in the right segment, as observed for the smaller dextrans in the experiments (see main text). The stationary state is reached faster in the approximate system compared to the actual measurements in the main text, due to the much smaller z-dimension. Perfect agreement between the numerical model and the analytical solution is obtained.
\clearpage

\section{Error Estimate for Numerical Analysis}\label{sec:errors}
\begin{figure*}
\includegraphics[]{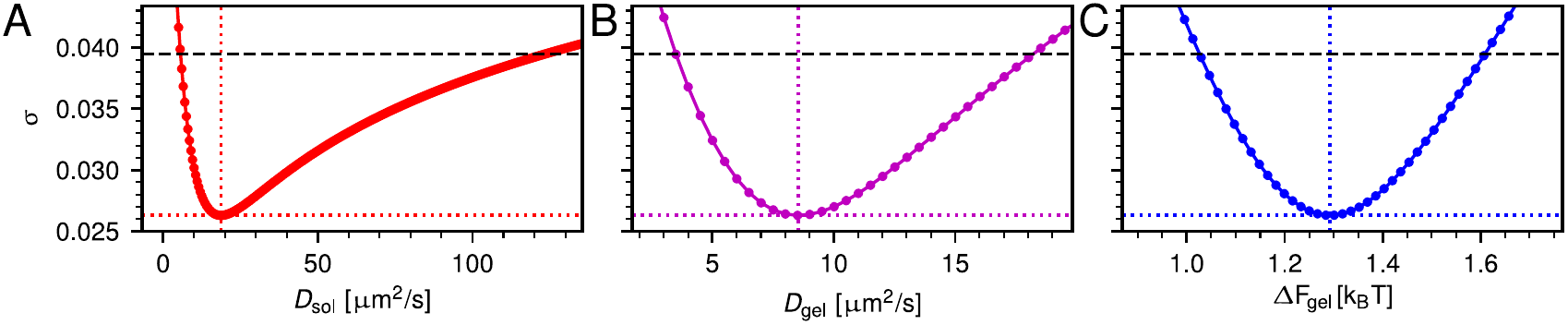}
\caption{Error estimation of the fitted values for $D_\text{sol}$ (A), $D_\text{gel}$ (B) and $\Delta F_\text{gel}$ (C) for measurements of $M_\text{dex}=40~$kDa dextran molecules diffusing into the \textit{hPG-G10} hydrogel. Fitted optimal values for the parameters are indicated by dotted lines, while a 50\% change in $\sigma$ is shown by the dashed black line.
A  larger value of the diffusion constants does not affect the agreement with the experimental data as strongly as a smaller value.}
\label{fig:error}
\end{figure*}

In order to determine confidence intervals for the fitted parameters of $D_\text{sol}$, $D_\text{gel}$ and $\Delta F_\text{gel}$, the values are varied from the optimum until the agreement with the experimental data is 50\% worse than for the optimal parameter values. Figure~\ref{fig:error} shows an exemplary analysis of the fitted parameters influence on the error. All parameters are varied independently, meaning that the error is always computed while keeping all other parameters fixed at their optimal values. Also, the fitted values for $d_\text{int}$ and $z_\text{int}$ are not changed but kept at their optimum. It is apparent that increasing the fitted diffusion constants does not affect the agreement with the experimental data as strongly as a decrease (see Figure~\ref{fig:error}A and B). Changing the free energy difference influences the numerical error $\sigma$ more symmetrically, meaning increasing $\Delta F_\text{gel}$ has the same influence on the error as decreasing it.

\clearpage

% BIBLIOGRAPHY
\bibliography{supp_refs}